\documentclass[prl,twocolumn,showpacs,preprintnumbers,amsmath,amssymb]{revtex4}

\usepackage{graphicx}
\usepackage{dcolumn}

\newcommand{\imag}[0]{\dot{\imath}}
\newcommand{\epszero}[0]{\varepsilon_0}
\newcommand{\epsinf}[0]{\varepsilon_{\infty}}
\newcommand{\epsbar}[0]{\bar{\varepsilon}} 

\newcommand{\beq}{\begin{equation}}
\newcommand{\eeq}{\end{equation}}
\newcommand{\beqa}{\begin{eqnarray}}
\newcommand{\eeqa}{\end{eqnarray}}

\begin{document}

\title{Wigner crystallization in a polarizable medium}

\author{G. Rastelli and S. Ciuchi}

\affiliation{Istituto Nazionale di Fisica della Materia and 
Dipartimento di Fisica\\
Universit\`a dell'Aquila,
via Vetoio, I-67010 Coppito-L'Aquila, Italy}

\begin{abstract}

We present a variational study of the 2D and 3D Wigner crystal phase of large
polarons. The method generalizes that introduced by 
S.\ Fratini,P.\ Qu{\'{e}}merais [Mod.\  Phys.\  Lett. B {\bf 12} 1003 (1998)]. 
We take into account the Wigner crystal normal modes rather than a single 
mean frequency in the minimization procedure of the variational free energy. 
We calculate the renormalized modes of the crystal as well as the
charge polarization correlation function and polaron radius. The
solid phase boundaries are determined via a Lindemann criterion, suitably generalized to 
take into account the classical-to-quantum cross-over.

In the weak electron-phonon coupling limit, the Wigner crystal parameters are
renormalized by the electron-phonon interaction leading to a stabilization of
the solid phase for low polarizability of the medium. Conversely,
at intermediate and strong coupling, the behavior of the system depends
strongly on the polarizability of the medium.

For weakly polarizable media, a density crossover occurs inside the solid
phase when the renormalized plasma frequency approaches the phonon
frequency.  At low density, we have a renormalized polaron Wigner crystal,
while at higher densities the electron-phonon interaction is weakened
irrespective of the {\it bare} electron-phonon coupling.

For strongly polarizable media, the system behaves as a
Lorentz lattice  of dipoles. The abrupt softening of the internal polaronic
frequency predicted by Fratini and Quemerais is observed near the actual
melting point only at very strong coupling, leading to a possible liquid
polaronic phase for a wider range of parameters.

\end{abstract}
\date{\today}
\maketitle

\section{Introduction} 

As it was first proposed by Wigner last century \cite{Wigner} the long
range Coulomb interaction is able to stabilize a crystal of electrons, which
eventually melts upon increasing the density at a quantum critical point. 
Experiments done on heterostructures \cite{2DWC} and quantum Monte Carlo
simulations confirm this scenario \cite{Ceperley_3D,Ceperley_2D}.  The presence
of impurities is known to stabilize the crystal phase in two dimensions 
\cite{Tanatar}. Another mechanism which could help the stabilization of the
crystal phase is the effect of a polar material. As a single electron moves in
a polar crystal, it  polarizes its environment creating a new quasi-particle: a
Fr\"ohlich large polaron, with an enlarged effective
mass \cite{Kuper,Devreese}. 
One expects then an enlargement of the Wigner crystal phase. 

Interesting properties of the liquid phase in polar doped semiconductors arise
also due to the interaction with the polarization,  as for example, the
mixing between  plasmons and longitudinal optical (LO)
phonons \cite{Mooradian}. Such a mixing can be explained by assuming a long
range interaction between the carriers and optical lattice vibrations  of the
Fr\"ohlich type \cite{Varga}.  The resulting coupled LO-phonon-plasmon modes
(CPPM´s) are found in polar semiconductors ({\sl n}-type GaP or {\sl p}-type
GaAs) \cite{Irmer}. Another interesting playground for this kind of 
physics is the surface-polaron, i.e. electron close to the surface of 
a polar crystal, which have been intensively studied especially for intermediate 
electron-phonon coupling $\alpha$ as in InSb where $\alpha \sim 4.5$ 
\cite{Sun_Gu}  
or in AgCl where $\alpha \sim 3$ \cite{Zhang_Xiao}. It has been also observed that 
the gate materials (SiO$_2$,Al$_2$ O$_3$ in organic thin films 
in field transistors are polar dielectrics and the interaction between 
the electrons and the surface phonons of the polar dielectric 
is relevant \cite{Kirova_Bussac}. 

The aim of this work is to study the stabilization of the Wigner Crystal phase
and its properties in the the presence of a polarizable medium. 
We consider a general model in which the key feature is the
presence of long range interaction which arise from direct Coulomb interactions
between electrons and from the polarizable medium.

The presence of long range interactions, high polarizability and low carrier
density is also a common feature of high-temperature superconductors. Of course, in
these materials, short range interactions and lattice effects play an important
role. Nonetheless,  polarons have been detected by optical measurements in the
antiferromagnetic insulating phase of both superconducting and parent cuprates
\cite{Kim,Taliani,Calvani}. Moreover, some evidence of strong electron-phonon
coupling effects has been given recently in the underdoped regime \cite{Lanzara}. 
A new interesting physics is introduced when studying these materials  by the
fact that the carrier concentration can be varied  from very low to
sufficiently high density.  
Prediction on optical properties and more specifically the
behavior of the so called Mid Infrared Band (MIR) by varying the doping has
been proposed according polaronic models \cite{Simone_1,Kastner,Lorenzana}  
as well as its interpretation as charge ordering in stripes \cite{Calvani_stripes}. 
A similar behavior has been also found in the optical 
properties of potassium  doped Barium Bismutate\cite{BaBiO_timusk}.

When we consider a system composed of many interacting large polarons we are
faced with the problem of screening of {\it both} electron-electron (e-e) and
electron-phonon (e-ph) interactions as we increase the carrier density. 
A density crossover is therefore expected when 
the doping concentration is varied  so that 
the plasma frequency 
approaches the  optical longitudinal phonon
frequency $\omega_{LO}$.

At high density,   phonons cannot follow the much faster plasma oscillations of
the  electron gas and therefore they do not contribute to the screening of the 
e-e interactions. On the other hand, the electronic density fluctuations 
screen  the e-ph interaction leading to the undressing of the  electrons from their
polarization clouds. As a consequence, the polaronic mass renormalization
is hugely reduced \cite{Mahan}. In this case, the plasma frequency of the pure  
electron gas $\omega^2_P= 4 \pi e^2 \rho / m$ is renormalized by the 
high frequency dielectric constant $\epsinf$
\begin{eqnarray} 
\label{eqn:omega_ren_high}  
\omega^2_{P,H}=
\frac{\omega^2_P}{\epsinf}  
\end{eqnarray}
In the high density region, the self energy has been studied by a perturbative approach 
for  weak  e-ph coupling ($m_{pol} \simeq m$) in the metallic phase\cite{Mahan}. 
The validity of this approach is ruled by the  condition of $\omega^2_{LO} \ll \omega^2_P/
\epsinf$, since  in this regime $\omega^2_P/ \epsinf$ is  representative of
electron density fluctuations.  In this case, the electron screening weakens the
{\sl effective} e-ph coupling constant and it is argued that  the perturbative
approach is suitable also for semiconductors which have intermediate values of the
{\it bare} e-ph  coupling in the low doping phase. The same
results have been obtained  at weak and intermediate couplings by a ground
state study \cite{Devreese_Lemmens}.  An approach which is able to span the
strong e-ph coupling regime has been presented in ref. \cite{Iadonisi_Capone}, where
an untrapping transition is found by increasing the density via the plasmon
screening of the e-ph interaction. There it is concluded that there is no
polaron formation at high density,  irrespectively of the strength of the bare
e-ph coupling constant.

At low density, the phonon energy scale (phonon-frequency) exceeds
the electronic energy scale (plasma frequency).  In this limit, the phonons can
follow the oscillations of the slower electrons and  they screen the e-e
interaction.  Thus, the frequency of the  electron collective modes  is 
renormalized by the static dielectric constant $\epszero$.  Moreover, in the case
of intermediate and strong e-ph coupling,  polarons are formed \cite{Bozovic} so
that the appropriate expression for the  general renormalized plasma frequency
becomes 
\begin{equation} 
\label{eqn:omega_ren_low}   
\omega^2_{P,L}
= \frac{m}{m_{pol}} \frac{\omega^2_P}{\epszero} 
\end{equation}  
where $m_{pol}$ is the polaron mass. In the case of $GaAs$ the 
mass renormalization  due  to
the e-ph interaction is negligible, and usually
 eq.(\ref{eqn:omega_ren_low}) is
used to interpret the experimental data with $m$, the band mass of the 
carriers, in place of $m_{pol}$ \cite{Irmer}.

In ref.\cite{DeFilippis_Cataudella} an approximation is developed which allows
to study  a system of many interacting large polarons in the intermediate/low density regime 
for  weak and intermediate coupling strengths. The phonon degrees of freedom are eliminated  
by a generalized Lee-Low-Pines transformation \cite{LeeLowPines} 
obtaining
an effective pair potential  between electrons which is non-retarded, with a short range 
{\sl attractive} term and a long range  Coulomb repulsive term, statically screened 
by $\epszero$. 
The role of the inverse polarizability parameter $\eta=\epsinf/\epszero$ 
is evident in ref. \cite{DeFilippis_Cataudella}. In the case of 
$\eta \ll 1$,which is hereafter reported as the high polarizability regime,  
repulsive interaction and the retarded
phonon-mediated attractive interaction are comparable leading to
a softening of the energy of the collective modes
at a finite value of
the wave vector $k$,  signaling a  charge density wave instability.
The attractive interaction term  between the electrons plays a crucial role
also at very low density where the ground state can be bi-polaronic 
below a certain value of the polarizability parameter 
\cite{Verbist_Smondyrev}, or can undergo a solid/liquid phase
transition  similar to the Wigner Crystallization (WC) \cite{Quemerais}. 
In ref. \cite{Simone_meanfield,Simone}, a Large
Polaron Crystal (LPC) is studied using a path-integral scheme.  
In ref \cite{Simone}, for $\eta =1/6$ (high polarizability regime),  
the authors conclude that in the weak and
intermediate e-ph coupling regimes at $T=0$ the LPC melts toward a polaron
liquid, but in the  strong  coupling regime a phonon instability appears near
the  melting. The authors argue this behavior from the softening of a long
wavelength collective mode due to the  e-e dipolar-interaction. 
A study which shows that the presence of long range order is not necessary for
this kind of scenario has been presented in ref. \cite{Lorenzana} for 
a simplified model of a classical liquid 
of interacting dipoles, which are the polarons treated {\sl \`a la } Feynman. 
The dipolar mode (internal frequency) is renormalized by the mean field 
of the other dipoles and it is shown to soften as the  density increases, 
leading to the dissociation of the dipole (polaron).

The present work generalizes  the approach of ref. \cite{Simone}  using a
formalism which allows to span from high ($\eta \ll 1$)  to low polarizability
regime ($\eta \simeq 1$). 
We calculate the boundaries of the solid phase in
three as well as in two dimensions. We also calculate,  within the solid phase, the
correlation function between the electron density  and the charge polarization
density. Our results confirm the relevant role of the  parameter $\eta$ in the
strong e-ph coupling regime. According to the values of this parameter two
distinct behaviors are found:

i) the high polarizability regime in which we found a scenario similar to that
of ref. \cite{Simone} i.e. the melting of the crystal is driven by the
instability of the internal polaronic mode. Interestingly our more quantitative
prediction push the instability-driven melting toward very strong couplings
leaving the possibility of a liquid polaronic phase for a wider range of
parameters.

ii) the low polarizability regime, studied here even at strong coupling, in
which we found that the undressing transition  argued
in the liquid phase \cite{Iadonisi_Capone} occurs also in the solid phase. 
Nonetheless e-ph interaction is able to stabilize the crystal against 
the liquid phase even for moderately polarizable mediums.

This paper is organized as follows: in the first section we illustrate the
model and the approximations used,  we introduce the quantities of interest,
and we also discuss  the Lindemann criterion used to determine the transition
temperature. In the second section,  we present the results in the three
dimensional case. In the third section, the results of the two dimensional case
are compared to the those in 3D. The conclusions are reported in the last
section. Appendices contain technical details of the calculations.

\section{ I - The model and the method} 

\subsection{a) The model}
The model describes a system of N interacting electrons  in a $D$-dimensional space, 
which are coupled to longitudinal (undispersed) optical phonons. 
The Hamiltonian of the model is a generalization of that 
introduced by Fr{\"o}hlich for a single large polaron \cite{Evrard} 
to N-large polarons \cite{Mahan}. 
We consider electrons as distinguishable particles. 
This approximation  is justified inside the solid phase,  
where the overlap between the wavefunctions 
of different localized electrons is negligible \cite{Carr}. 
Using the Path Integral technique \cite{Feynman_Hibbs} phonons can be easily 
traced out 
taking advantage of their gaussian nature and we end up with the 
following partition function \cite{Simone_meanfield}:
\begin{equation}  \label{eqn:Z_eff} 
\mathcal{Z} = 
\oint \prod_{\imath}
\mathcal{D} \left[\vec{r}_{\imath}(\tau) \right] 
e^{- \frac{1}{\hbar} \mathcal{S}_{eff}}
\end{equation}
where $\oint$ means the functional integration over all cyclic space-time paths 
of the particles $\vec{r}_{\imath}(\tau)$ between zero and 
$\beta = \hbar / k_B T$. 
The effective electron action reads 
\begin{equation} \label{eqn:S_eff}
\mathcal{S}_{eff}  =  
\mathcal{S}_{K}  + \mathcal{S}_{e-e} + 
\mathcal{S}^{self}_{e-ph-e}  + \mathcal{S}^{dist}_{e-ph-e} + \mathcal{S}_{J}
\end{equation}
where
\begin{eqnarray} \label{eqn:S_eff1}
\mathcal{S}_{K} &= & 
\int_0^{\beta} \!\!\!\! d\tau
\sum_{\imath} \frac{1}{2} m {\left| \dot{\vec{r}}_{\imath} (\tau) \right|}^2  \\
\mathcal{S}_{e-e} & = &  \frac{e^2}{2 \epsinf}
\int_0^{\beta} \!\!\!\! d\tau 
\sum_{\imath \neq \jmath} \frac{1}{\left| \vec{r}_{\imath}(\tau)-\vec{r}_{\jmath}(\tau)  \right|}  
\label{eqn:S_dist_e-e} \\
\mathcal{S}^{self}_{e-ph-e}  & = & - \frac{\omega_{LO} (1-\eta) e^2}{4 \epsinf} 
\int_0^{\beta} \!\!\!\! d\tau \!\! \int_0^{\beta} \!\!\!\! d\sigma 
\sum_{\imath} 
\frac{D_o \left( \tau-\sigma \right) }
{\left| \vec{r}_{\imath} (\tau)-\vec{r}_{\imath} (\sigma) \right|}  \nonumber \\
& & \label{eqn:S_self_e-ph-e} \\
\mathcal{S}^{dist}_{e-ph-e}  & = & - \frac{\omega_{LO} (1-\eta) e^2}{4 \epsinf} 
\int_0^{\beta} \!\!\!\! d\tau \!\! \int_0^{\beta} \!\!\!\! d\sigma 
\sum_{\imath \neq \jmath} 
\frac{D_o \left( \tau-\sigma \right) }
{\left| \vec{r}_{\imath} (\tau)-\vec{r}_{\jmath} (\sigma) \right|} \nonumber \\
& &  \label{eqn:S_dist_e-ph-e} \\
\mathcal{S}_{J}  & = &  
\beta \frac{{\left(e \rho_J\right)}^2}{2 \epszero} V \!\! \int \!\!\! \frac{d\vec{r}}{r}
\! - \!  
\int_0^{\beta} \!\!\!\! d\tau 
\sum_{\imath} \!\! 
\int \!\!\! d\vec{r}
\frac{e^2 \rho_{J} / \epszero} {\left| \vec{r}_{\imath}(\tau) - \vec{r} \right|}  \label{eqn:S_e-J} 
\end{eqnarray}
Here  $e^2$ is the electron charge, $m$ is 
the electron band mass
and $V$ is the
volume. $(e \rho_J)$ is the static jellium charge density.
The integration of phonons leads to the appearance of retarded 
e-e interaction terms --eqs.(\ref{eqn:S_self_e-ph-e},\ref{eqn:S_dist_e-ph-e})--, 
where the phonon propagator is
\begin{equation} \label{eqn:propagator_o}
D_o (\tau) = \frac{\cosh \left( \omega_{LO}[\beta/2 - \tau]\right)}
{ \sinh \left( \beta \omega_{LO} /2 \right)}
\end{equation}
Using polaronic units (p.u.) ($\hbar \omega_{LO}$  for 
 energy, $1 / \omega_{LO}$ for  imaginary time $\tau$ and  
$\sqrt{\hbar / m \omega_{LO}}$ for  lengths) $\mathcal{S}_{e-ph-e}$ becomes 
proportional to the dimensionless e-ph coupling constant $\alpha$  defined as 
\begin{equation} \label{eqn:alpha}
\alpha =  
\frac{e^2}{\sqrt{2}} \frac{1 - \eta}{\epsinf} \sqrt{ \frac{m}{\hbar^3 \omega_{LO} } } 
\end{equation}
while $\mathcal{S}_{e-e}$ will be proportional to the e-e 
coupling constant
\begin{equation} \label{eqn:alpha_e}
\alpha_e = \frac{\sqrt{2} e^2}{\epsinf} \sqrt{ \frac{m}{\hbar^3 \omega_{LO} } } 
\end{equation}
the ratio $ \alpha_e / \alpha = 2 / \left(1 - \eta \right)$ is thus solely 
determined by $\eta = \epsinf / \epszero$ :  
when $\eta \simeq 1$ the Coulomb repulsion overwhelms   
the attraction mediated by phonons, while they
become  comparable for $\eta \ll 1$.
Therefore, in the Fr\"ohlich
model,  the inverse polarizability parameter rules  the relative weight 
between the repulsive and attractive (phonon-mediated)  interactions. 
This attraction can lead to a bi-polaronic ground state 
as $\alpha > \alpha_c(\eta)$ 
\cite{Verbist_Smondyrev}. 
Roughly speaking this condition implies strong couplings $\alpha > \alpha_c$
{\it and} high polarizability $\eta < \eta_c$ where $\alpha_c = 9.3$ 
and $\eta_c = 0.131$ in 3D, $\alpha_c = 4.5$ and $\eta_c = 0.158$ 
in 2D case \cite{Verbist_Smondyrev}. 
We have investigated the system for two values of $\eta$,  representative
respectively of the high and low polarizability regimes, and several values 
of the e-ph coupling $\alpha$.  We choose respectively
$\eta= 1/6$ as in ref.\cite{Simone}, which gives $\alpha_e / \alpha = 2.4$, and 
$\eta= 0.90519$  so that the coupling  $\alpha_e / \alpha$ is increased by a
factor of ten \cite{footnote_epsinf}. 
For these values of $\eta$, no bipolaron ground state exists.

\subsection{b) The harmonic variational approximation in the solid phase}

We generalize the harmonic variational approach originally introduced in  
ref.\cite{Simone_meanfield} to study the model eq.(\ref{eqn:S_eff}).
First of all we recall here the variational theory in the path integral formalism.
Let us consider a suitable trial action $\mathcal{S}_T$ which depends 
on some variational parameters. Substituting $\mathcal{S}_{eff}$ with $\mathcal{S}_{T}$ 
in  eq.(\ref{eqn:Z_eff}) we obtain  the partition function  $\mathcal{Z}_T$ 
for the trial action and the free energy associated to it  
$\mathcal{F}_T = - k_B T \ln \mathcal{Z}_T$. 
Then the exact free energy can be expressed as  
\begin{equation} \label{eqn:F}
\mathcal{F} =
\mathcal{F}_T - k_B T \ln 
{\left< e^{- \frac{1}{\hbar} \Delta \mathcal{S} }\right>}_T 
\end{equation}
where $\Delta \mathcal{S} = \mathcal{S}_{eff} - \mathcal{S}_{T}$ 
and the mean value ${\left<  \dots \right>}_T$ is 
\begin{equation} \label{eqn:mean}
{\left< \dots \right>}_T = \frac{1}{\mathcal{Z}_T}
\oint 
\prod_{\imath}
\mathcal{D} \left[\vec{r}_{\imath}(\tau) \right] 
\left( \dots \right) 
e^{- \frac{1}{\hbar} \mathcal{S}_{T}}
\end{equation}
The variational free energy is obtained by a cumulant expansion of the 
logarithm appearing in eq. (\ref{eqn:F}). At first order in $\Delta \mathcal{S}$ 
it reads:
\begin{equation} \label{eqn:F_var}
\mathcal{F}_{V}  =  
\mathcal{F}_T + \frac{1}{\beta}  {\left< \Delta \mathcal{S} \right>}_T 
\end{equation}
where $\mathcal{F}_{V} \geq \mathcal{F}$. 
To define a suitable trial action we proceed in two steps as in 
ref.\cite{Simone_meanfield}.
First we treat the self interaction term $\mathcal{S}^{self}_{e-ph-e}$ 
of eq. (\ref{eqn:S_self_e-ph-e}) {\it a la Feynman} \cite{Feynman,Schultz}. 
Therefore we substitute $\mathcal{S}^{self}_{e-ph-e}$  
with $\mathcal{S}_{Feyn}$  
\begin{equation} \label{eqn:S_Fey} 
\mathcal{S}_{Feyn}  =  
\!\!\! 
\frac{(v^2 - w^2) m w}{8} \sum_{\imath} \!\!
\int_0^{\beta} \!\!\!\!\! d\tau \!\! \int_0^{\beta} \!\!\!\!\! d\sigma
 D_{V} (\tau-\sigma) 
{\left|\vec{r}_{\imath}(\tau)-\vec{r}_{\imath}(\sigma) \right|}^{2} 
\end{equation}
$v$ and $w$ are the two variational parameters. The variational propagator $D_V (\tau)$ 
is given by eq. (\ref {eqn:propagator_o}) with $w$ replacing $\omega_{LO}$. 
We remind that $S_{Feyn}$ eq.(\ref{eqn:S_Fey}) can be obtained by integrating out an 
action where each electron interacts elastically ($K_T~=~m \left( v^2 - w^2 \right)$)
with a fictitious particle of mass  $M_T~=~m \left[ (v^2/w^2) - 1 \right]$.
Then $v$ is the internal frequency and $1/\mu= 1/m + 1/M_T$ is the reduced mass 
of the two particle system

As a second step, we treat the $\mathcal{S}_{e-e}$, $\mathcal{S}_{J}$ 
eqs.(\ref{eqn:S_dist_e-e},\ref{eqn:S_e-J}) and 
the distinct part $(\mathcal{S}^{dist}_{e-ph-e})$ eq.(\ref{eqn:S_dist_e-ph-e})
of $\mathcal{S}_{eff}$ in eq.(\ref{eqn:S_eff}) by means of a harmonic approximation. 
Expressing the position of the electrons around the Wigner lattice points as 
$\vec{r}_{\imath} = \vec{u}_{\imath} + \vec{X}_{\imath}$ where
$\vec{X}_{\imath}$ are the vectors of the Bravais lattice
(b.c.c. in 3D, hexagonal in 2D) and
omitting the constant terms of the solid phase potential energy, 
we obtain the following 
harmonic variational action:
\begin{eqnarray}   \label{eq:Svar} 
\mathcal{S}_{T}  &=&
\mathcal{S}_{K} 
+ \mathcal{S}_{Feyn} 
+ \mathcal{S}^{H}_{J} + \mathcal{S}^{H}_{e-e} 
+ \mathcal{S}^{H,dist}_{e-ph-e} \label{eqn:S_T_Harm} 
\end{eqnarray}
where 
\begin{eqnarray}
\mathcal{S}_K & = & 
\int_0^{\beta} \!\!\!\! d \tau  \sum_{\imath} \frac{1}{2} m
{|\dot{\vec{u}}_{\imath}(\tau)|}^2  \label{eq:SK} \\
\mathcal{S}^{H}_{e-J} +  \mathcal{S}^{H}_{e-e}  &=&
\int_0^{\beta} \!\!\!\! d\tau 
\sum_{\imath} \frac{1}{2} m \frac{\omega^2_W}{\epszero}  {|\vec{u}_{\imath}(\tau)|}^2 \nonumber  \\
&+&  \int_0^{\beta} \!\!\!\! d\tau  \frac{e^2}{2 \epsinf}  \sum_{\imath \neq \jmath} 
\vec{u}_{\jmath}(\tau) \overline{\mathcal{I}_{\imath \jmath}} 
\vec{u}_{\imath}(\tau) \label{eqn:S_H_dist_e-e}
\end{eqnarray}
\begin{equation} \label{eqn:S_H_dist_e-ph-e}
\mathcal{S}^{H,dist}_{e-ph-e} = 
- \frac{\omega_{LO} e^2}{4 \epsbar} \sum_{\imath \neq \jmath} 
\int_0^{\beta} \!\!\!\! d\tau \!\!  \int_0^{\beta} \!\!\!\! d\sigma 
 D_o(\tau-\sigma)  
\vec{u}_{\jmath}(\sigma) \overline{\mathcal{I}_{\imath \jmath}}
\vec{u}_{\imath}(\tau)  
\end{equation}
In eq.(\ref{eqn:S_H_dist_e-e}), the Wigner frequency is defined as usual in 3D  as 
$\omega^2_{W,3D} = \omega^2_P / 3$ (for the 2D case see eq.(\ref{eqn:w_Wig_2D}) in Appendix B). 
The force constants  ${\left[ \overline{\mathcal{I}_{\imath \jmath}} \right]}_{\alpha \beta}$ are
obtained through a harmonic  expansion
for the Coulomb potential (see appendix B). 
In our calculations, we neglect the anharmonic terms in  $\Delta \mathcal{S}$  
of eq. (\ref{eqn:F_var}), therefore we get   
\begin{eqnarray} \label{eqn:F_harm_var}
\mathcal{F}_{V} & = & 
\mathcal{F}_{T} + \frac{1}{\beta} {\left< \mathcal{S}^{self}_{e-ph-e} -\mathcal{S}_{Feyn} \right>}_T 
\end{eqnarray}

We have minimized $\mathcal{F}_{V} / N$  varying   $w,v$ at given density 
and temperature keeping $\alpha$ and $\eta$ fixed. Minimization is constrained by a 
convergence condition on the  gaussian integrals appearing in $\mathcal{F}_{V}$.
The constrained minimization procedure is described in appendix C. 

So far, the discussed scheme appears very similar to the one of 
ref.\cite{Simone_meanfield}.  However, we stress that the $\mathcal{F}_{V}$,
which we have minimized  to obtain the variational parameter $v$ and $w$ ,
contains the hetero-interaction terms $\mathcal{S}^{H}_{e-e}$   and
$\mathcal{S}^{H,dist}_{e-ph-e}$, which are not included in the minimization 
procedure of ref. \cite{Simone_meanfield}.
Moreover, we have also used the {\sl whole} trial action $\mathcal{S}_{T}$ 
eq.(\ref{eq:Svar}) to calculate the mean electronic fluctuation  
which we have used in the Lindemann rule, as explained in the following section.

\subsection{c) Lindemann rule and phase diagrams}

To determine the solid-liquid transition we use the 
phenomenological Lindemann criterion, suitably generalized to 
take into account the classical-to-quantum cross-over 
\cite{4_Hansen_Mokovotich}:
\begin{equation} \label{eqn:del_L}
\frac{ {\left< {\left| \vec{u} \right|}^2 \right>}_{eff} }{d^2_{n.n.}}  
= \gamma^2  \left( \eta_q \right)
\end{equation}
in the l.h.s. of eq. (\ref{eqn:del_L}) we have the 
 Lindemann ratio between 
the mean fluctuation of the electrons around its 
equilibrium position and the nearest neighbors distance $d_{n.n.}$.
When it exceeds a critical value (r.h.s. of eq. (\ref{eqn:del_L})),
the solid melts. 
In eq.(\ref{eqn:del_L}) ${\left< \dots \right>}_{eff}$ is the 
average taken over $\mathcal{S}_{eff}$ eq.(\ref{eqn:S_eff}). The 
average is carried out at the zeroth order in the 
cumulant expansion as an average over $\mathcal{S}_T$ eq.(\ref{eq:Svar}).

Contrary to the classical liquid-solid transition, where 
the Lindemann rule predicts the full melting line using a constant 
$\gamma = \gamma_{cl}$, in the case of a quantum crystal 
an interpolating formula for $\gamma$ is necessary to determine the 
melting line as obtained by comparing the free-energies of the two 
phases calculated using quantum simulations \cite{Ceperley_3D}.
Hence the analytic expansion of the quantum corrections to the classical free 
energy respect to the quantum parameter $\eta_q$  and the 
zero-temperature melting density provides the interpolating function 
(r.h.s. of eq.\ref{eqn:del_L}) for $\gamma \left( \eta_q \right)$
\cite{4_Hansen_Mokovotich}. 
$\eta_q$ is defined for the pure electron gas as the ratio between zero point 
and thermal activation energies as:
\begin{eqnarray} \label{eqn:eta_q}
\eta_q &=& \frac{\hbar \omega_p}{2 k_B T}.
\end{eqnarray}

We have chosen for the function $\gamma(\eta_q )$ the form of refs.
\cite{Nagara,Ceperley}:
\begin{eqnarray} \label{eqn:gamma}
\gamma(T,r_s) & = & \gamma_{q} - 
\frac{\gamma_{q} - \gamma_{cl}}{1 + A \eta^2_q} 
\end{eqnarray}

Formula (\ref{eqn:gamma}) has a single interpolation parameter $A$ 
which we take as $A=1.62~\cdot~10^{-2}$ in 3D \cite{Nagara} and 
$A=3~\cdot~10^{-2}$ in 2D \cite{Ceperley}. 

The chosen value of $\gamma_{cl} = 0.155$ is such that the classical
transition lines ($T=2 / \Gamma_c r_s$  a.u.) are recovered in both the 3D
($\Gamma_c=172$ from ref. \cite{Nagara}) and 2D ($\Gamma_c=135$  from ref.
\cite{Ceperley}) cases.
The value $\gamma_{q}=0.28$ is chosen to reproduce the zero temperature
quantum transition in 3D ($r_s =100$ a.u. from ref. \cite{Nagara} ) and 2D ($r_s =
37$ a.u. from ref. \cite{Ceperley}).

Roughly speaking, the transition curve is limited  by 
the classical line $T = \left(2 / \Gamma_c\right) 1/r_s$ and 
the quantum melting $1/r_s = 1 / r^c$. The actual transition 
curve is a smooth interpolation between these two limiting behaviors.
Of course, the precise knowledge of the interpolation formula 
(i.e. the knowledge of parameters appearing in it) 
is critical only for the determination of the transition line at high
temperatures (see fig.\ref{fig:phasediagram}). 

We notice that the particular values of the parameters entering 
in eq.(\ref{eqn:gamma}) depend on the kind of statistics 
(boson, fermion) and on the system parameters only via the ratio $\eta_q$ \cite{footnote_gamma}. 
This parameter depends on the mass of the particles via
$\omega_p$, which measures the zero point energy of the oscillator which eventually melts \cite{nota_m_pol}.
Therefore, to generalize the Lindemann criterion to the interacting large
polaron system we are left with the alternative of choosing between the electron and the polaron 
effective mass in eqs. (\ref{eqn:eta_q},\ref{eqn:gamma}).

The polaron exists as a well defined quasiparticle  when {\it both} $k_B T \ll \hbar \omega_{LO}$
\cite{CiuchiPierleoni} and $\hbar \bar{\omega}_P \ll \hbar \omega_{LO}$. 
The second condition relies on the effectiveness of the e-ph interaction, as explained
in the introduction. Therefore, if both conditions are fulfilled, 
we have to replace $\omega_P$ in eq. (\ref{eqn:eta_q}) by  
$\bar{\omega}_P$ given by eq. (\ref{eqn:omega_ren_low}). 
In this case, between the classical  $(\eta_q \simeq 0)$ and 
and quantum melting $(\eta_q \rightarrow \infty)$, we have a polaronic Wigner Crystal.
This is the case of the high polarizability  $(\eta~=~0.17)$.

For low polarizability  $(\eta~=~0.9)$, a cross-over 
occurs inside the solid phase 
when $\hbar \bar{\omega}_P \sim \hbar \omega_{LO}$ and the coupling is intermediate or strong, 
as it will be discussed in details later on. 
In this case, we still have a classical melting of polaronic quasi-particles, 
but the quantum melting involves the undressed electrons. 
In the classical regime 
(low density), the transition line does not depend appreciably 
on the quantum parameter, as
$\gamma$ attains its classical limit 
($\eta_q \rightarrow 0$).
In the quantum regime at high density and low temperatures
($\eta_q \rightarrow \infty$) the function 
$\gamma$ eq. (\ref{eqn:gamma}) saturates to its quantum value $\gamma_{q}$and 
the density $r_c$  of the
quantum melting does not depend of the choice for the quantum parameter
$\eta_q$. Instead, a pronounced dependency on the actual value of the quantum
parameter is expected in the calculation of the melting line at 
high temperatures and intermediate densities.

For $\eta=0.9$ 
we choose the high density estimate $\omega_P / \sqrt{\epsinf}$ 
as the plasma frequency entering  
in eq. (\ref{eqn:eta_q}).
This choice produces, in the intermediate temperature/density 
region, an upward deviation (fig. \ref{fig:phasediagram} lower panel) from 
the classical slope. This is a drawback of our approximation, which 
is however correct at low temperatures for both low and high density.

We finally discuss to which extent we use the Lindemann criterion in 2D, 
and more generally on the applicability of the harmonic theory in 2D.
This is related to the well known problem of  the existence of two dimensional 
crystalline long-range order at finite temperature \cite{Mermin}.
In a pure electron gas, for $T=0$, this problem does not arise and the properties of the
system in the harmonic approximation have been studied extensively 
\cite{Bonsall,Maradudin}.
The general statement for the classical impossibility 
of 2D crystalline long-range order 
was first pointed out by Peierls \cite{Peierls}.
Landau \cite{Landau} gave a general argument according to which fluctuations
destroy crystalline order possessing only a one or two dimensional periodicity.
The first microscopic treatment of the problem
(not valid in case of Coulomb interaction) is due to Mermin \cite{Mermin}:
his proof is based on Bogolyubov's inequality that leads to the 
conclusions that the Fourier component of the mean density  
is zero for every vector $k$ in the thermodynamic limit.
Motivated by the interest of the 2D electron gas,  Mermin's proof  
was critically re-examined for the long range potential \cite{Baus,Alastuey} .
We discuss here the argument of Peierls  for the 2D electron crystal. 
The mean square thermal fluctuations of a generic classical particle
diverges in two dimensions for an infinite harmonic crystal.
At low density, we have $\eta_q \simeq 0$ 
and the mean electronic  fluctuation can be  
approximated by the classical value
\begin{eqnarray}  
\left< u^2 \right>_{Cl,WC} &=&   \frac{D k_B T}{2 m \omega^2_P}
\mathcal{M}_{-2} \label{eqn:u_classic} \\ 
\mathcal{M}_{-2} &=& \int \!\!\! d \omega \rho \left(
\omega \right)  \frac{\omega^2_P}{\omega^2} \label{eqn:M_inv_2}
\end{eqnarray}  
where $\mathcal{M}_{-2}$ is the dimensionless second inverse moment of the
density of the states (DOS) of charge fluctuation normal modes in the pure WC 
($\rho (\omega)$). 
Since long-wavelength acoustical vibrational modes scale  as
$\omega = c_s k$,  the DOS is given at low energies by 
$\rho (\omega) \sim \omega$ for $\omega \rightarrow 0$ \cite{Crandall} 
and the integral eq.(\ref{eqn:u_classic}) diverges logarithmically.
However, a lower cut-off in the frequency spectrum, 
which exists for a large but finite system 
studied in laboratory \cite{Crandall} or in a computer simulation 
\cite{Ceperley_2D,Gann}, removes 
the logarithmic  divergence. 
We have chosen a cut-off frequency 
which corresponds to a fixed number of particles $N \simeq 5 \cdot 10^{5}$.
The dependence of the cut-off is discussed in appendix A.
There and later on it is shown that our results are cut-off independent for
low temperatures and  density near the quantum critical point.
Therefore we will discuss 2D case only in this region.

\subsection{d) Correlation functions and polaron radius}

We now introduce the correlation functions between the 
electron and the polarization densities for a system with N electrons, and a
measure of the polaron radius.

The  polarization density vector of the medium 
is associated to the optical phonon modes 
$Q_{\vec{q}}$ through the relation \cite{Devreese}:

\begin{equation} \label{eqn:P}
\vec{P} \left( \vec{r} \right) =   
\sum_{\vec{k}} 
\imag
\frac{\omega_{LO}}{\sqrt{4 \pi \epsbar  V}}
\frac{\vec{k}}{\left| k \right|}
e^{\imag \vec{k} \vec{r}} Q_{\vec{k}}.
\end{equation} 
The induced charge density is defined by \cite{Devreese}
\begin{equation} \label{eqn:n_i}
n_{i}\left( \vec{r} \right) 
= - \frac{1}{e}
\vec{\nabla} \cdot \vec{P}\left( \vec{r} \right)
\end{equation} 
Correlation between a given electron and the induced charge density can be defined as:
\begin{equation} \label{eqn:n_imath}
C_1 \left( \vec{r'} ,\vec{r} \right)  = 
\frac{ \left< \rho_1 (\vec{r}) n_{\imath} (\vec{r'})\right>}
{\left< \rho_1 (\vec{r}) \right>}  
\end{equation} 
with $\rho_{1} (\vec{r}) =  \delta \left( \vec{r} - \vec{r}_{1} \right)$.
In eq.(\ref{eqn:n_imath}) we have chosen the appropriate normalization for 
the correlation function between {\sl one} electron and the 
polarization. Integrating out the phonons we arrive at the following expression, 
in which we express all quantities in terms of averages weighted by
the effective action  eqs.(\ref{eqn:S_eff})
\begin{equation} \label{eqn:rhodef}
C_1 \left( \vec{r'} ,\vec{r} \right)
= \!\!
\frac{1}{\epsbar} \int^{\beta}_{0} \!\!\!\! 
d \tau \frac{\omega_{LO}}{2} D_o(\tau)
\frac{
{\left< \rho_1 (\vec{r}) 
\rho (\vec{r'},\tau) \right>}_{eff}
}
{{\left< \rho_1 (\vec{r}) \right>}_{eff}}
\end{equation}
In eq.(\ref{eqn:rhodef}) $\rho (\vec{r}',\tau)$ is the path 
density defined by:
\begin{equation} \label{eqn:def-ro-self}
\rho (\vec{r'},\tau) =
\delta \left( \vec{r'} - \vec{r}_1 (\tau) \right) +
\sum_{\imath \neq 1}
\delta \left( \vec{r'} - \vec{r}_{\imath} (\tau) \right) 
\end{equation}
where we have explicitly separated the contribution $\rho_1(\vec{r'},\tau)$ 
due to the electron $1$ from
the remainder.
The first contribution in r.h.s. of eq. (\ref{eqn:def-ro-self}) 
give rise to a self term in the correlation function 
eq. (\ref{eqn:rhodef}) given by
\begin{eqnarray} \label{eqn:C_self}
C^{self}_1 
&=& 
\frac{1}{\epsbar} 
\int^{\beta}_{0} \!\!\! d \tau \frac{\omega_{LO}}{2} D_o(\tau)
\frac{
{\left< \rho_1 (\vec{r}) 
\rho_1 (\vec{r'},\tau) \right>}_{eff}
}
{{\left< \rho_1 (\vec{r}) \right>}_{eff}} 
\end{eqnarray}
Notice that in the limit of a single isolated polaron, 
this correlation function reduces to the one evaluated in ref. 
\cite{CiuchiPierleoni}.
Assuming an electron at origin $(\vec{r}=0)$, we have 
$C^{self}_1$ depending only on $\vec{r'}$.
The radial induced charge density $g(r)$ can be defined as
\begin{equation} \label{eqn:g_r}
g (r) =  
r^{D-1}
\int d^D \Omega \;\;\; 
C^{self}_1  \left( \vec{r} \right) 
\end{equation}

Using this function, we can define, as a measure of the polaronic  radius, 
the square root of the second moment of $g(r)$ 
\begin{equation} \label{eqn:R_p}
R_p = 
{\left(
\int^{\infty}_0  \!\!\! d r \; r^2  g(r)
\right)}^{1/2}
\end{equation}

The actual calculation for the mean values appearing in eq.(\ref{eqn:C_self}) 
are carried out at the zeroth order of the variational cumulant expansion. 
Explicit calculations are reported in Appendix E.

\section{II - Results in 3D} 

Here we compare the low  $(\eta=0.9)$ and high $(\eta=1/6)$ polarizability cases in 3D. 
\begin{figure}[htbp]
\begin{flushleft}
\includegraphics[scale=.33,angle=270.]{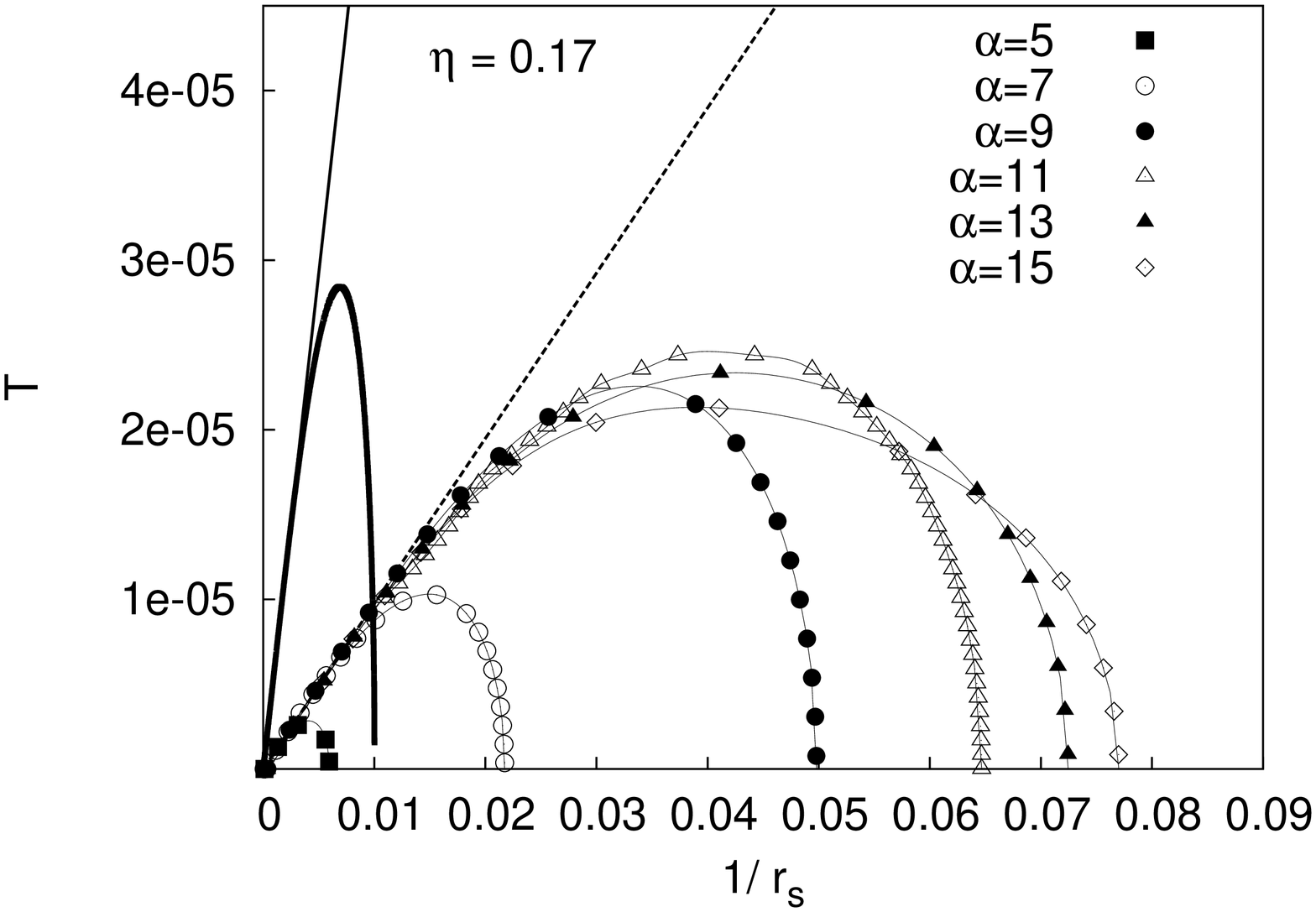}
\includegraphics[scale=.33,angle=270.]{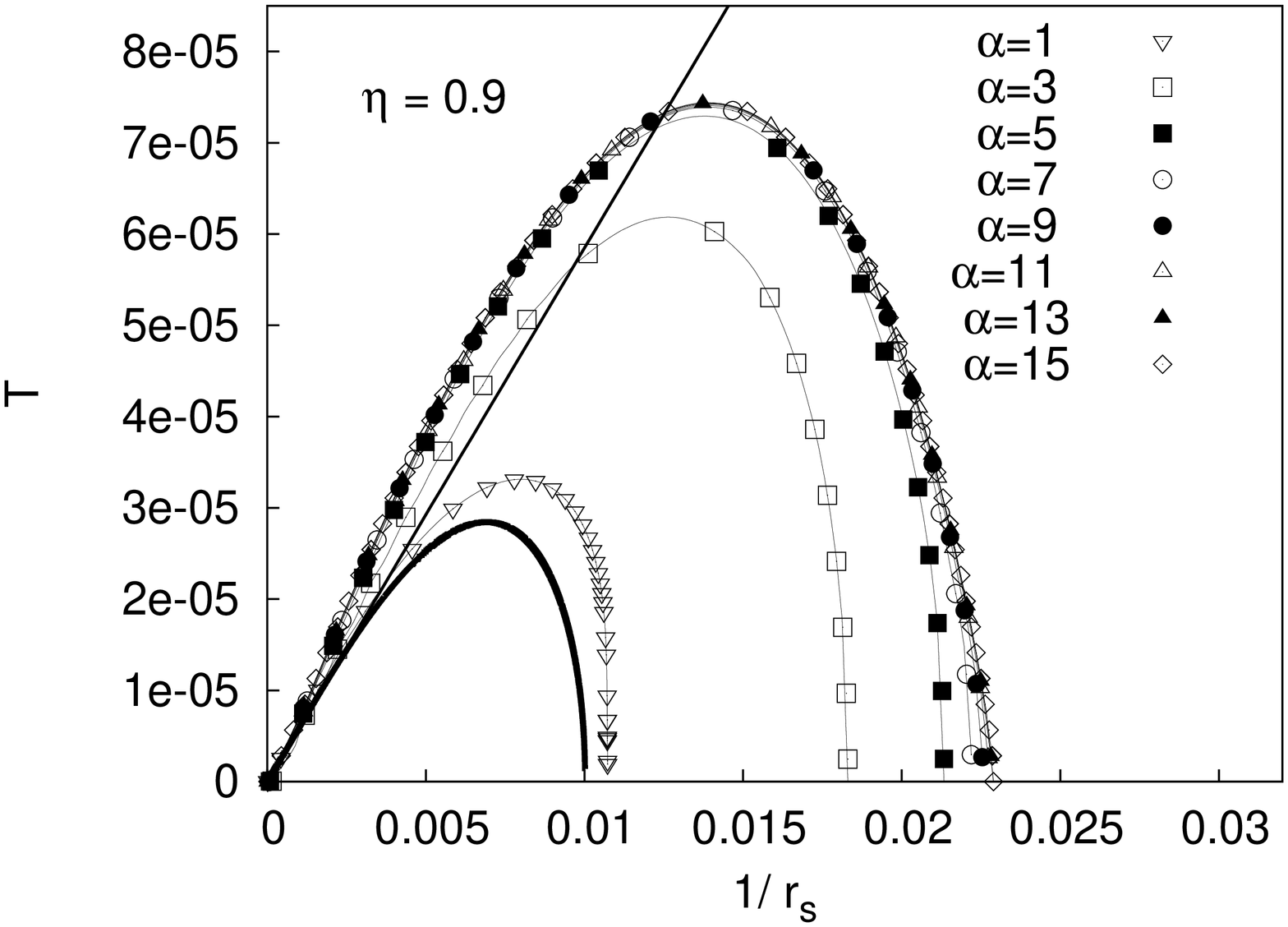}
\caption{
Phase diagrams for a 3D LPC for $\eta=0.17$ (upper panel) 
and $\eta=0.9$ (lower panel).
Atomic units (a.u.) are used for temperature and $r_s$ (see text).
Solid phase is enclosed below transition lines.
In both the upper and the lower panels 
continuous bold curve is the pure WC transition line and
solid line gives the classical melting.
In the upper panel dashed line is the renormalized classical melting.}
\label{fig:phasediagram}
\end{flushleft}
\end{figure}
For each polarizability, the electron-phonon 
coupling constant $\alpha$  spans from weak to  strong coupling regime: 
$\alpha= 1,3,5,7,9,11,13,15$. 
Phase diagrams obtained through the Lindemann criterion are shown in figs. 
\ref{fig:phasediagram} where the solid-liquid transition lines of the LPC are compared 
to that of the pure Wigner crystal. 
Density is expressed in term of the adimensional parameter $r^3_s = a^3_o / [(4 \pi / 3) \rho]$, 
where $a_o$ is the Bohr radius with $(m=m_e,\epsinf=1)$. 
A common feature of both the low and high polarizability cases
is the enlargement {\sl in density scale} of the solid phase 
as far as e-ph coupling increases. 
However, in both cases, the solid phase cannot be stabilized for any density by
increasing the e-ph interaction, and the quantum melting point saturates at a
maximum value when the e-ph coupling is very strong. 
To illustrate this different behavior it is worth to 
introduce a simplified model.

\subsection{a) A simplified model}

In the simplified model, introduced in ref. \cite{Simone}, 
the electrons interact with each other and with {\sl all} 
the fictitious particles $(\{ \vec{R}_{\imath} \})$ with mass $M_T$  
which represent the polarization of the medium.  
After integration of the fictitious particles, we obtain the effective 
electronic lagrangian $\mathcal{L}_{eq}$. The effective harmonic 
lagrangian $\mathcal{L}_{eq}$  generated by the simplified model 
corresponds exactly to the lagrangian of the action 
$\mathcal{S}_{T}$ eq.(\ref{eq:Svar}) with the parameter $w=\omega_{LO}$. 
This approximation restricts the space of variational parameters, 
and therefore gives rise to a worse estimate for the free energy. 
Nonetheless, it allows to describe the physics of the system in
a simplified fashion. 

Each WC's branch  is  splitted in two branches for the LPC and
the frequencies of the system are given 
by the two roots $\Omega^2_{\pm} (\omega_{s,\vec{k}})$ (eqs.(24,25) 
of the work \cite{Simone}), where $\omega_{s,\vec{k}}$ are 
the WC frequencies with wave vector
$\vec{k}$ and branch index $s$. 
The expression for the mean fluctuation ${\left< u^2 \right>}_{eq}$ 
of electrons around their equilibrium value in the simplified model is easily 
obtained by inserting $w = \omega_{LO}$ in the variational expression 
${< u^2 >}_T$ 
(see Appendix D, eqs.\ref{eqn:sigma2},\ref{eqn:del2_+_eq},\ref{eqn:del2_-_eq}). 
The $\Omega_{\pm}$ branches give rise to a natural splitting of
contributions to the fluctuation 
\begin{equation} \label{eqn:del2_eq} 
\frac{{\left< u^2 \right>}_{eq}}{d^2_{n.n.}} =
\frac{{\left< u^2 \right>}_{+}}{d^2_{n.n.}}  + 
\frac{{\left< u^2 \right>}_{-}}{d^2_{n.n.}}
\end{equation}

In the low density regime of the simplified model \cite{Simone}, 
i.e. when phonons are much faster than density fluctuations,
the spectrum can be decomposed into the renormalized  WC frequencies 
$\tilde{\Omega}_{-} (\omega_{s,\vec{k}})$
and the polaronic optical frequencies 
$\tilde{\Omega}_{+} (\omega_{s,\vec{k}})$, which 
can be obtained by expanding the general solutions 
$\Omega^2_{\pm} (\omega_{s,\vec{k}})$  with respect 
to the parameter $\epsilon_{s,\vec{k}}$ 
defined as 
\begin{equation} \label{eqn:eps_s_k}
\epsilon_{s,\vec{k}} = \omega^2_{s,\vec{k}} / \left( \epszero v^2 \right)
\end{equation}
which is small for {\sl all} frequencies $\omega_{k,s}$ of WC normal modes  
at low density  regime.
 
The first part of the spectrum represents the {\sl low} frequencies associated to the 
oscillation of the center of mass $(m_{pol} = m + M_T)$ 
of the two-particle system, i.e. 
the electron and  its relative fictitious particle (polarization), 
while the second part of the spectrum  describes the dipolar modes associated to the 
internal motion of oscillating electron-fictitious particle system (fig.1 of ref.\cite{Simone}).  
Dipolar modes are weakly dispersed around the frequency $\omega_{pol}$ 
(eq.(25) of ref.\cite{Simone}) defined as the $k=0$ mode 
of the polaronic branches. 
It represents the internal frequency of oscillation of
the electron inside its polarization well.

\subsection{b) Classical and renormalized quantum melting}

Now let us consider the classical transition. 
This transition is located in the low density regime of the 
simplified model. Using the low density expansion for 
the spectrum  
$( \tilde{\Omega}_{-} (\omega_{s,\vec{k}}),\tilde{\Omega}_{+} (\omega_{s,\vec{k}}) )$, 
it is possible to associate each term, 
${\left< u^2 \right>}_{+}$ and ${\left< u^2 \right>}_{-}$,
to a definite degree of freedom of the two-particle system, i.e. 
to the fluctuation of the center of mass 
\begin{equation}  \label{eqn:del2_-low_eq}
{\left< u^2 \right>}_{-} \simeq 
\int \! d \omega \; \rho \left( \omega \right)
\frac{\hbar D
\coth \left[
\hbar \left( \sqrt{\frac{m}{\epszero m_{pol}}} \right) \omega/ 2 k_B T 
 \right] }
{2 \; m_{pol} \; \left( \sqrt{\frac{m}{\epszero m_{pol}}} \right) \omega } 
\end{equation}
and to the fluctuation associated to the internal dipolar mode with  
$\vec{\rho} = \vec{u} - \vec{R}_T$ and reduced mass $\mu$ 
\begin{equation} \label{eqn:del2_+low_eq}
{\left< u^2 \right>}_{+} \simeq {\left( \frac{M_T}{m + M_T}\right)}^2
\frac{\hbar D}{2 \mu \omega_{pol}}
\coth \left( \frac{\hbar \omega_{pol}}{2 k_B T}
\right)  
\end{equation}

In this case we can easily estimate the ratio between electronic fluctuations in LPC and 
in WC by taking into account only the renormalized WC spectrum, i.e. the 
fluctuation associated to the center of mass eq.(\ref{eqn:del2_-low_eq}). 
Using eq.(\ref{eqn:u_classic}) we have $<u^2>_{LPC}/<u^2>_{WC}=1/\epsilon_0$
then by Lindemann criterion eq. (\ref{eqn:del_L}) and by 
eq. (\ref{eqn:u_classic}) at a given
density, the critical temperature ratio also equals $1/\epsilon_0$
\begin{equation}
\label{eq:ratioclass}
\frac{ T^{Cl}_{LPC} }{ T^{Cl}_{WC} } = 
\frac{1}{\epszero}. 
\end{equation} 
Therefore, the slope of the classical transition line is
lowered by the same factor, as can be seen in fig. \ref{fig:phasediagram} (upper
panel), where $\epszero$ is appreciably large. 

The quantum melting is ruled by the zero point fluctuations of the electronic 
oscillations. A zero temperature estimate for the pure WC gives
\begin{eqnarray} 
\label{eqn:QuantMelt}
\left< u^2 \right>_{WC}
&=&   \frac{\hbar D}{2 m \omega_P} \mathcal{M}_{-1} \\
\mathcal{M}_{-1} &=& \int \!\!\! d \omega \rho \left( \omega \right) 
\frac{\omega_P}{\omega}
\end{eqnarray} 
where $\mathcal{M}_{-1}$ is the dimensionless inverse moment of the WC DOS. If
we consider only the renormalized WC spectrum eq.(\ref{eqn:del2_-low_eq}),  
and we take into account eq.(\ref{eqn:QuantMelt}), we get for the LPC
\begin{equation} \label{eq:ratioquant}
\frac{ \left< u^2 \right>_{Q,LPC} } {\left< u^2 \right>_{Q,WC}} 
= \left ( \frac{m\epszero}{m_{pol}} \right )^{1/2} {\left( \frac{r_s}{r_s(WC)} \right)}^{3/2}.
\end{equation} 
then using Lindemann criterion we obtain at the quantum critical point
($r_s=r_c$)
\begin{equation}
\label{eqn:rc_scale}
\frac{r_c(WC)}{r_c}=\frac{m_{pol}}{m\epszero}.
\end{equation}
eq.(\ref{eqn:rc_scale}) generalizes the result of the 
ref.\cite{Simone_meanfield} where the Lindemann rule was 
discussed within a mean field approach.  

\begin{figure}[htbp]
\begin{center}
\includegraphics[scale=.4,angle=0.]{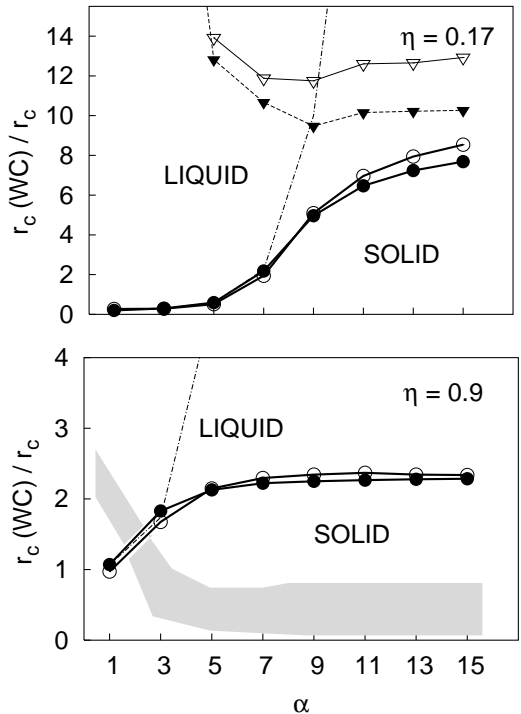}
\caption{Zero temperature phase diagram in the 2D (open symbols) and 
3D (solid
symbols) cases. In 2D $\alpha$ has been scaled according the zero density limit.
Circles are  
the scaled quantum melting $r_c$ {\sl vs} e-ph coupling constant $\alpha$.
Dashed line is the renormalized 
quantum melting transition curve from eq.(\ref{eqn:rc_scale}).
Upper panel: $\eta=0.17$.  
Triangles locates the softening of $\omega_{pol}$. 
Lower panel: $\eta=0.9$.  The shaded area   
encloses  the cross-over region inside the solid phase.}
\label{fig:3D2D_summary}
\end{center}
\end{figure}

At high polarizability  ${\left< u^2 \right>}_{-}$
eq.(\ref{eqn:del2_-low_eq}) is the leading term in the mean
electronic fluctuation ${\left< u^2 \right>}_{eq}$ eq.(\ref{eqn:del2_eq}) 
near the quantum melting for small and intermediate couplings $\alpha \le 7$. 
In this case, the quantum melting density scales as eq.(\ref{eqn:rc_scale}).
Notice that at weak coupling the mass renormalization is weak, 
but phonon screening  through $\epszero$ dominates,
leading to quantum melting at lower densities than in a purely 
electronic Wigner Crystal 
(upper panel figs.\ref{fig:phasediagram},\ref{fig:3D2D_summary}). 
At low polarizability eq.(\ref{eqn:rc_scale}) 
is valid up to $\alpha\simeq 3$ (fig.\ref{fig:3D2D_summary}).  

Upon increasing the coupling, $m_{pol}$ scales as $\sim \alpha^4$ 
in strong coupling and eq.(\ref{eqn:rc_scale}) predicts a divergence 
of the quantum melting density. As shown in  figs. \ref{fig:phasediagram},\ref{fig:3D2D_summary},
the quantum melting density saturates to an $\alpha$-independent value at strong coupling, 
and the prediction of eq. (\ref{eqn:rc_scale}) is no longer valid.
We will see in the next subsection  that deviation from the prediction of eq.
(\ref{eqn:rc_scale}) arise from different reasons in low and high
polarizability cases.

\subsection{c) High polarizability: softening of internal mode}

This is the case in which the polarization gives a large contribution to the
total interaction energy of the system. The system can be thought as being
composed by interacting dipoles which are made by electrons surrounded by their
polarization. 

\begin{figure}[htbp]
\begin{center}
\includegraphics[scale=.4,angle=270.]{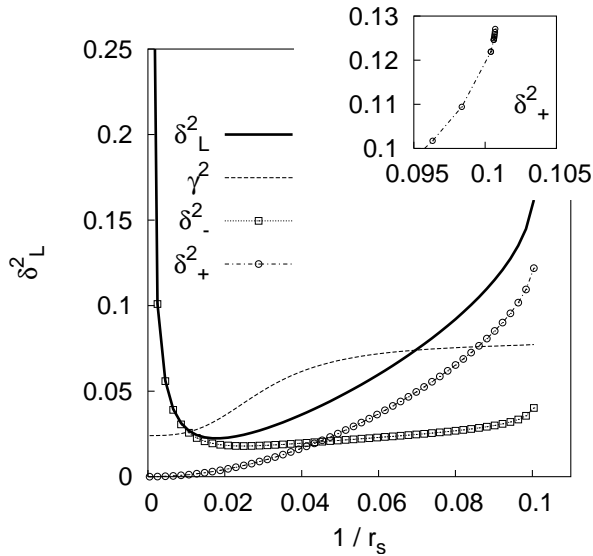}
\caption{
The Lindemann ratio (solid line) and the function $\gamma$ (dashed line)
for $\eta=0.17$ and $\alpha=13$, $T=1. 10^{-5}$ (a.u.). 
Contributions $\delta^2_\pm=<u^2>_\pm/ d^2_{n.n.}$ of the 
simplified model eq.(\ref{eqn:del2_eq}) are also
shown. The inset shows the abrupt slope increase of the term $\delta^2_+$.
}
\label{fig:fluctuations_02_a9}
\end{center}
\end{figure}

In the case of strong  e-ph coupling 
we observe a saturation of the critical quantum melting
density. In  fig. \ref{fig:fluctuations_02_a9} the electronic  fluctuation
is reported for $\alpha=13$. Contrary to the small/intermediate coupling
case, ${\left< u^2 \right>}_{+}$  eq.(\ref{eqn:del2_+low_eq})
is now the leading term near the quantum melting.
The melting density given by  eq.(\ref{eqn:rc_scale}) is not a good estimate
due to the contribution of polaronic optical modes which is now 
important at the quantum melting. The same scenario of
ref.\cite{Simone} is recovered: the optical polaronic frequencies
drive the melting at strong coupling and high polarizability. 
Moreover we notice that ${\left< u^2 \right>}_{+} \sim (1 / \omega_{pol})$ and,
as density approaches a critical value, $\omega_{pol}$ softens inducing 
an abrupt increase of  electron fluctuation  which is dominated by the
term  ${\left< u^2 \right>}_{+}$ (see fig.\ref{fig:fluctuations_02_a9}).  
Same behavior for $\omega_{pol}$ is reported in ref.\cite{Simone} 
and explained in term of the attractive  interaction between the polarons 
(polarization catastrophe). 
We stress however that employing a more quantitative Lindemann
criterion together with a self-consistent variational calculation of all
Feynmans' parameters we get quantum melting in a region in which $\omega_{pol}$
do not actually soften. 
As a result the softening of internal polaronic frequency approaches quantum melting only
asymptotically for very large $\alpha$ (fig.\ref{fig:3D2D_summary}).. 

\begin{figure}[htbp]
\begin{center}
\includegraphics[scale=.4,angle=0.]{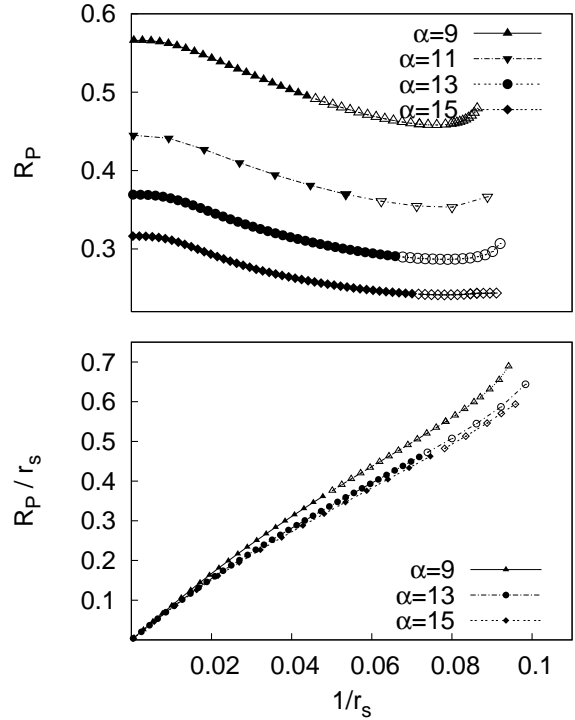}
\caption{
Polaron radius in polaronic units (upper panel) and polaron radius scaled with  $r_s$
(lower  panel) {\it vs} $(1 / r_s)$ (a.u.) 
for different $\alpha$ and $\eta=0.17$
at low temperature ($T=5 \; 10^{-3}$ p.u.).
Filled points refer to the solid phase.}
\label{fig:radius_02}
\end{center}
\end{figure}

Saturation occurs to value of $r_c$ which seems to lie  in the high density
regime where our approach could be  questionable. We must stress however that 
in the pure electron gas, the parameter $r_s$ is a measure of  {\it both}
coupling and density. Indeed $r_s$ can be obtained  from the ratio of the Fermi
energy to the mean Coulomb interaction,   even if  scaled with the band mass 
and the dielectric constant of the host medium, which  are anyway fixed, i.e.
no-density dependent. 
\begin{table}[htpb]
\begin{tabular}{||p{1cm}||p{2.25cm}|p{2.25cm}||}
\hline
  $\alpha$    & $r_c$  &  $r^{*}_c$ \\
\hline
\hline
  1     & 510  &   99.8 \\
\hline
  3     & 334  &   99.3 \\
\hline
  5     & 168  &   99.5 \\	
\hline
  7     &  46  &   99.8 \\
\hline
  9     &  20 &   197.5 \\
\hline
 11     &  15 &   452.2 \\
\hline 
\end{tabular} 
\caption{The critical value at the melting of the density ($r_c$) 
and coupling ($r^*_c$) 
parameters as function of $\alpha$ for high polarizability $\eta=0.16$.}
\label{tab:r_c}
\end{table}
If we introduce another coupling  in the system, as the e-ph interaction,   the
two concepts are distinct.  Global interaction is not only a function of the 
density but it is also a function of the e-ph coupling $\alpha$.  Now in the
high polarizability case polarons are well defined  as quasi-particles and we
can use $m_{pol}(\alpha)$  as effective mass while the repulsive interactions
is  screened by $\epszero$ in the low density regime.  Only in this case we can
introduce a  measure of the {\it coupling} through the  parameter $r^{*}_s =
(m_{pol} / \epszero m ) r_s$.   For the low polarizability case the last
assumption is not valid   as explained onward. The values of $r^{*}_s$ at the
quantum melting ($r^{*}_c$)  are reported in tab.\ref{tab:r_c}.  When the
coupling $\alpha < 7$ the quantum melting can be extimated thru 
eq.(\ref{eqn:rc_scale})  which means  $r^{*}_c \simeq r_c(WC) = 100$ that is 
the {\it coupling} parameter $r^*_c$ tends to the value  of the Wigner crystal
melting of a 3D electron-gas. On the contrary in the strong e-ph coupling  the
values of the effective coupling parameter $r^{*}_c$ are much bigger than those
of the density $r_c$ due to the huge enhancement of the polaron mass. 

Of course the exchange effects {\it at} the crystal melting  are relevant and
they can be taken into account only phenomenologically  in our harmonic
approximation (see footnote \cite{footnote_gamma}  and discussion in sec.
II.C).  However in the solid phase we must notice that these effects are
ruled in LPC by the parameter $r^*_s$ rather than $r_s$  making them
much more negligible that those at the same density in the pure electron gas. 
To realize this fact we assume that the localized electronic wave function is a
gaussian of variance $\sigma$ then the overlap between two of these functions
at distance $r_s$ is proportional to $\exp(-r^2_s/4\sigma^2)$. Now $\sigma$ in
the harmonic approximation can be extimated as  $\sigma^2=1/2 m_{pol}\omega_W$
where $\omega^2_W=\omega^2_{P,L}/3$ is the LPC Wigner frequency and 
$\omega^2_{P,L}$ is given by eq. (\ref{eqn:omega_ren_low}). Then is immediate
to see that $r^2_s/4\sigma^2=\sqrt{r^*_s}/2$ a results that can be compared
with the same for electron-gas \cite{Carr} in which appears $r_s$ and a
different coefficient due to a more elaborate variational procedure. Taking
into account data of table \ref{tab:r_c}  we have that exchange effects are
{\sl a fortiori}  negligible in a first approximation in the case of the strong
e-ph coupling where the quantum melting occurs at huge coupling parameter
$r^{*}_c$.

In fig. \ref{fig:radius_02} we show the behavior of the polaron radius as a
function of density. While in the solid phase it remains almost constant, when
approaching the melting density it suddenly increases. This behavior can be
understood by taking into account that the polaron radius is essentially
determined by the diffusion in imaginary time of the electron path defined  in
eq.(\ref{eqn:d_tau}) of Appendix C  (see also
eqs.\ref{eqn:R_p_T},\ref{eqn:defll} in Appendix E). Its maximum value occurs at
$\tau = \beta/2$ which  diverges at the softening of the polaronic frequency
$(\omega_{pol} \sim 0)$.  Polaronic clouds tends to overlap
(fig.\ref{fig:radius_02} lower panel). However, the polaronic nature of each
particle  of the LPC is preserved up to quantum melting.

\subsection{d) Low polarizability: cross-over in solid phase }

In this regime $(\eta \sim 1)$, the repulsive interactions among electrons overwhelm the
attractive interactions due to the polarizability of the background, as can be seen by
the relative weight of e-e and e-ph interaction coupling constant
eqs.(\ref{eqn:alpha},\ref{eqn:alpha_e}).  However, self-trapping effects are still
present at least at strong  coupling and at low density, where electrons are localized.

Eq.(\ref{eqn:rc_scale})  quantitatively describes quantum melting   in
the low polarizability case only at weak coupling $(\alpha \leq 3)$.
When  $\alpha$ exceeds this value, 
a cross-over between a polaronic and a non polaronic phase is found 
inside the solid phase and the estimate of eq. (\ref{eqn:rc_scale}) 
no longer describes quantum melting. 
 
The low density regime, introduced in the previous subsection, is found 
only for the classical part of the crystal phase, where 
the polarization follows adiabatically the electron and 
the solid phase is a Wigner crystal made of polarons with an effective mass determined
by the e-ph interaction, in the way discussed for high polarizability.

\begin{figure}[htbp]
\begin{center}
\includegraphics[scale=.35,angle=270.]{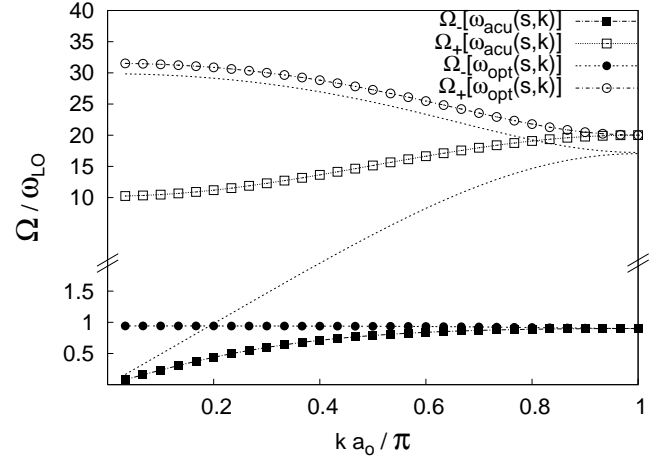}
\caption{
Frequencies of the system in the simplified model
as a function of $\vec{k}$ along the direction $(100)$,
for $\alpha = 5$ and $\eta=0.9$, at  $1/r_s = 2. 10^{-2}$
and at $T = 1.8 \; 10^{-5}$ (a.u.). 
Density is close to the quantum melting. 
$(\Omega_{-}[\omega_{acu}],\Omega_{+}[\omega_{acu}])$
result from the  splitting of $\omega_{acu}(s,k)$
the acoustical WC branch. 
$(\Omega_{-}[\omega_{opt}],\Omega_{+}[\omega_{opt}])$ result  
from the splitting of the high frequency WC optical branch
(eqs.(\ref{eqn:Om2_-high},\ref{eqn:Om2_+high})).
For comparison are shown the pure WC frequencies (dotted lines).
}
\label{fig:frequency_high}
\end{center}
\end{figure}

As far as the density increases (inside the solid phase), we observe that 
the two energy scale, $\omega_{LO}$ of phonons and $\omega_{P,L}$ and  
eq.(\ref{eqn:omega_ren_low}) of the renormalized WC frequencies, come close and  
we found a cross-over region {\sl inside the solid phase}, 
where the electrons and the polarization modes are mixed as
in the liquid phase CPPM (fig.1 of ref.\cite{Varga} and fig.1 of ref.\cite{Irmer}). 
Example of the general situation is given in fig.\ref{fig:frequency_high}.

\begin{figure}[htbp]
\begin{flushleft}
\includegraphics[scale=.35,angle=270.]{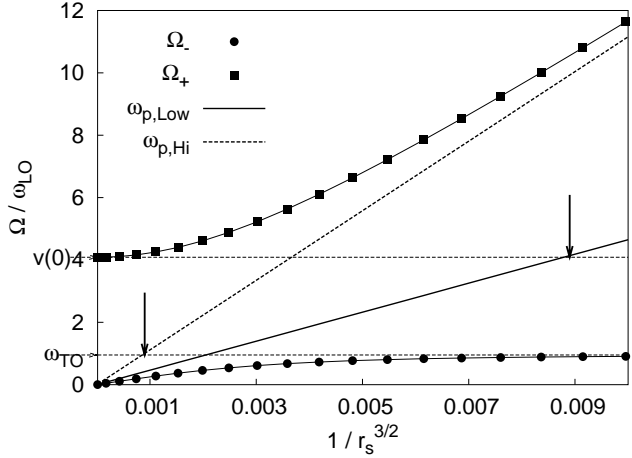}
\caption{Filled points are the typical frequencies of the simplified model
obtained with $\omega _{s,k}=\omega_P$ for $\alpha = 5$ and $\eta=0.9$.
Solid line is the 
low density renormalized plasma frequency eq.(\ref{eqn:omega_ren_low}).
Dashed line the high density renormalized plasma  frequency 
eq.(\ref{eqn:omega_ren_high}).
Arrows mark the crossover region (see text).
}
\label{fig:ren_plasma_frequency}
\end{flushleft}
\end{figure}

To estimate the density dependence 
of the LPC frequencies $\Omega_{\pm}(\omega_{s,k})$, let us substitute
$\omega_{s,\vec{k}}$ by the plasma frequency $\omega_P$. 
Results are reported in
fig.\ref{fig:ren_plasma_frequency}, which illustrates the density cross-over.
In the low density limit $(r_s \rightarrow \infty)$   $\Omega_{-} = \omega_{P,L}$
eq.(\ref{eqn:omega_ren_low}) while $\Omega_+$ converges to $\omega_{pol} \simeq v$, 
the  internal frequency for an single polaron.
In this case, the electrons are far apart, and the
\textquotedblleft~external~\textquotedblright
harmonic field  generated by the surrounding electrons of the crystalline array is weak
$(K_e \sim e^2/r^3_s)$.
Therefore the frequency of electron oscillation $(\omega^2_P \sim K_e /m)$
can be lower than that of the phonon $(\omega_{LO})$, and the
polarization   follows the electron oscillation.
The polaron vibrates as a whole with a lower frequency
$K_e / m_{pol} \sim {\omega}^2_P \left( m / m_{pol} \epszero \right)$.
The polarization charge
distribution is undisturbed as a first approximation, so that the value of the 
internal polaronic
frequency ($\sim v$) 
of an electron inside its polarization well doesn't change . 

By increasing the density, we approach  the opposite limit of strong
external field. Now the frequency associated to this field
is too large and the polarization cannot follow
the electron oscillation, so that each electron becomes  undressed from its polarization
cloud. In this case $\Omega_+$ approaches $\omega_P / \sqrt{\epsinf}$,  the high
density renormalized plasma frequency eq.(\ref{eqn:omega_ren_high}),  while
$\Omega_{-} \simeq \sqrt{\epsinf /\epszero}  \omega_{LO} = \omega_{TO}$  is the
characteristic  renormalized frequency of the polarization. 
We notice that at low density $\Omega_-$ 
gives a measure of the frequency of carrier density fluctuations, while in
the opposite limit of high density, the same role is played by $\Omega_+$.

As we can see from fig.\ref{fig:ren_plasma_frequency}, the  cross-over amplitude 
is determined by the conditions $\omega_{P,H} \simeq \omega_{TO}$ and
$\omega_{P,L} \simeq v$. 
It is interesting to compare our fig.\ref{fig:ren_plasma_frequency}
with the figure 1 of ref.  \cite{Varga}. We notice that the asymptotic boundary 
given there by the phonon frequency $\omega_{LO}$ 
 here it is played by the internal frequency. 

The cross-over of the renormalization of the plasma frequency  from low to high
density regime doesn't imply the melting of the crystal. Indeed, it is observed
within the boundary of  the solid phase estimated by  Lindemann criterion. This 
behavior is even
more clear once we consider the fluctuations of the  position of the electrons which
enter in the Lindemann criterion. 

The leading term for the Lindemann ratio at the classical melting  is  
$\delta^2_{-}={\left< {\left| u \right|}^2 \right> }_-/ d^2_{n.n.}$  which is
associated to the fluctuation of the center of mass eq.(\ref{eqn:del2_-low_eq}). Of
course, in the classical region quantum fluctuations are ineffective, the 
electrons and  its polarization  cloud behave as a single classical particle
with mass $m_{pol}$. The term   $\delta^2_{+}={\left< {\left| u \right|}^2
\right> }_+/ d^2_{n.n.}$ associated to the internal polaronic frequencies 
eq.(\ref{eqn:del2_+low_eq}) is indeed negligible.

To analyze the high density region where we meet
eventually the Lindemann criterion for quantum melting,
we notice that the condition $\epsilon_{s,\vec{k}} \gg 1 $,  
where $\epsilon_{s,\vec{k}}$ is defined in eq. (\ref{eqn:eps_s_k}),
can be fulfilled by the
majority of normal modes at high density. Of course, long wavelength
acoustical and even \textquotedblleft~optical~\textquotedblright
modes  in 2D have vanishing energies, but their spectral
weight is low enough to be neglected in the following considerations.
Expanding $\Omega_{\pm}(\omega_{s,k})$ in $1/\epsilon_{s,\vec{k}}$ we get
\begin{eqnarray}                
\Omega_{-}  & \simeq & \sqrt{\frac{\epsinf}{\epszero}} \omega_{LO}  \label{eqn:Om2_-high} \\
\Omega_{+}  & \simeq & \frac{\omega_{s,\vec{k}}}  { \sqrt{\epsinf}} \label{eqn:Om2_+high}
\end{eqnarray}
In  fig.\ref{fig:frequency_high}  
the general solutions  $\Omega_{\pm}(\omega_{s,k})$  are shown  for all the branches 
of the simplified model near the quantum melting.

The branches $(\Omega_{-}[\omega_{opt}],\Omega_{+}[\omega_{opt}])$ 
which results as splitting of optical model of the
Wigner crystal  $\omega_{opt}(s,k)$ are well described by
approximation of eqs.(\ref{eqn:Om2_-high},\ref{eqn:Om2_+high}).

The frequency dispersions $(\Omega_{-}[\omega_{acu}],\Omega_{+}[\omega_{acu}])$ 
of the modes  which originate from the splitting
of acoustical branches of the Wigner crystal is also reported. 
While at short wavelength, the
dispersion approaches the estimates given in  eqs.(\ref{eqn:Om2_-high},\ref{eqn:Om2_+high}) 
the long wavelength part of the spectrum is conversely described by the low density 
expansion $\tilde{\Omega}_{\pm}$.

Thus we have that at the quantum melting the low energy part of the spectrum
still behaves as in the low density regime. The modes depicted in the lower
part of  fig.\ref{fig:frequency_high} belongs to this part of the spectrum.

A measure of the wave-vector below which we have this behavior can be obtained by the
condition  $\epsilon_{s,\vec{k}}=1$. The associated energy scale is given by
$\omega^2_{c} = m \omega^2_{LO} / \left( \epszero  m_{pol} \right)$.
Contrary to low density regime
eqs.(\ref{eqn:del2_-low_eq},\ref{eqn:del2_+low_eq}), it is not possible to associate to each
term of the fluctuation  eqs.(\ref{eqn:del2_+_eq},\ref{eqn:del2_-_eq})  a
definite degree of freedom. However, expanding the electron fluctuation   with respect to the
parameter $\epsilon_{s,\vec{k}}$ for the frequencies  $\omega_{s,k} < \omega_c$ and  with
respect to  the parameter $1 / \epsilon_{s,\vec{k}}$ for the frequencies 
$\omega_{s,k} > \omega_c$ and using eq. (\ref{eqn:del2_eq})
the electron position fluctuations can be  approximated by:

\begin{eqnarray}
{\left< u^2 \right>}_{-} &\simeq&
\int^{\omega_c}_{0} \!\!\!\!\! d \omega \rho \left( \omega \right)
\frac{\hbar D
\coth \left[  \hbar 
\left( \sqrt{\frac{m}{m_{pol} \epszero}} \right) \omega / 2 k_B T  \right] }
{2 m_{pol} \left( \sqrt{\frac{m}{m_{pol} \epszero}} \right) \omega} \label{eqn:del2_-high_eq}\\
{\left< u^2 \right>}_{+} &\simeq&
{\left( \frac{M_T}{m + M_T}\right)}^2
\frac{\hbar D}{2 \mu \omega_{pol}}
\coth \left( \frac{\hbar \omega_{pol}}{2 k_B T}\right)
\int^{\omega_c}_{0} \!\!\!\!\! d \omega \rho \left( \omega \right)
 \nonumber \\
&+&
\;\;
\int_{\omega_c}^{\infty} \!\!\!\!\! d \omega \rho \left( \omega \right)
\frac{\hbar D}{2 m \frac{\omega}{\sqrt{\epsinf}} }
\coth
\left(
\hbar \frac{\omega}{\sqrt{\epsinf}} / 2 k_B T
\right)
\label{eqn:del2_+high_eq}
\end{eqnarray}

Notice that the interpretation of the fluctuations associated to electronic
motion in this case is different from that valid at low density. In particular
the high energy contribution (the second term of eq. (\ref{eqn:del2_+high_eq})
represents a Wigner crystal-like fluctuation with a low energy cutoff.
This is the largest contribution to the fluctuation at quantum melting
 and does not depend on e-ph interaction.

Indeed the leading term of fluctuations
at quantum melting  is ${\left< u^2 \right>}_{+}$. This is due to the vanishing
of the spectral weight associated to the low frequencies  $\omega <
\omega_c $ at high density (eq.(\ref{eqn:del2_-high_eq})).

The saturation of the quantum melting point can be seen in the phase diagram
of fig. \ref{fig:phasediagram} (lower panel). Two comments are needed.
First, in the case of very low e-ph coupling, the
density crossover does not occur inside the solid phase. Therefore, these arguments
do not apply. The quantum melting point depends on the e-ph coupling as we have
discussed in the previous section.
However a saturation of the quantum melting
density is observed clearly in fig.\ref{fig:phasediagram} for intermediate and
strong coupling.
As a second point we have to emphasize that the quantum melting
density {\it is not} that of a purely electronic Wigner crystal.

This fact can be explained by writing the total electron fluctuation as  the
sum of the two terms  $<u^2>_{Hi}$ and $<u^2>_{Low}$
where $<u^2>_{Hi}$ is  the contribution to fluctuations of modes having energies higher
(lower) than $\omega_c$. We notice from eq (\ref{eqn:del2_+high_eq}) that in both
LPC and WC case $<u^2>_{Hi}$ are the same. But while in the WC case
the  two terms are of the same order $<u^2>_{Low} \simeq  <u^2>_{Hi}$,
in the LPC case $<u^2>_{Low}  \ll <u^2>_{Hi}$ 
as far as the density increases. This is due to the renormalization 
of the low energy frequencies.
Therefore, the  electronic fluctuation in the LPC increases more slowly
with density than those of the WC. At given density, the electronic fluctuation of the WC 
is greater than those of the LPC and  this fact explains the shifting of the quantum melting
toward higher densities.

\begin{figure}[htbp]
\begin{center}
\includegraphics[scale=.4,angle=0.]{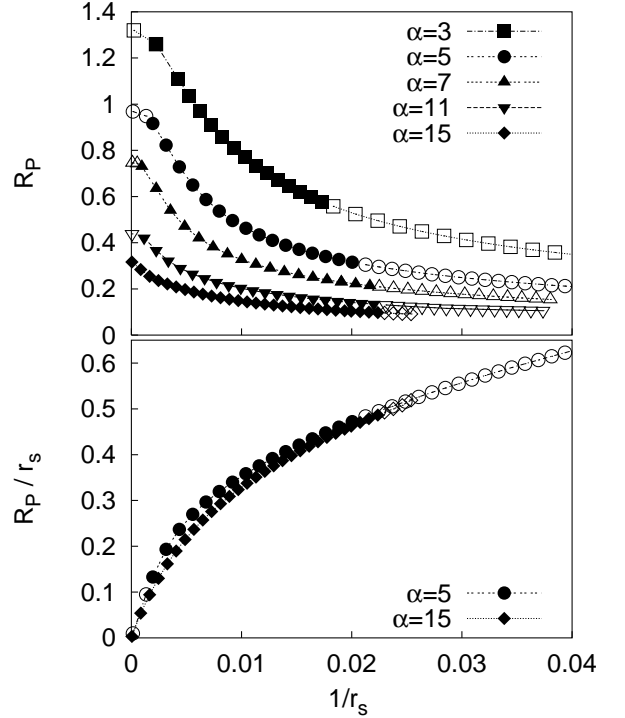}
\caption{
Polaron radius in polaronic units (upper panel) and polaron radius scaled with  $r_s$
(lower  panel) {\it vs} $(1 / r_s)$ (a.u.) 
for different $\alpha$ in the case $\eta=0.9$
at low temperature ($T =5 \; 10^{-2}$ p.u.).
Filled points refers to the solid phase.
}
\label{fig:radius_09}
\end{center}
\end{figure}

The cross-over is also evident in the polaron radius.
In the upper panel of fig. \ref{fig:radius_09}  we plot the  polaron radius
as defined by the eq.(\ref{eqn:R_p}). We see that for any value of the e-ph
coupling, the polaron radius tends to decrease as far as the density is increased.
We recall that as far as the renormalized plasma frequency  eq.(\ref{eqn:omega_ren_low}) 
exceeds the phonon frequency, we enter in a region in which the polarization
is adiabatically slow compared to the electronic motion. Therefore, the electronic
charge appears as a static distribution whose radius decreases upon increasing
the density and the polaron radius follows this trend. The crossover is evident
by scaling the polaron radius with  $r_s$, as  reported in the lower panel of
fig. \ref{fig:radius_09} at intermediate and strong $\alpha$. Notice that as in
the high polarizability case at the transition $R_p/r_s \simeq 0.475$.

It is possible to estimate the high density limit of the radial distribution of
the induced charge. Using the condition $\omega_o \ll \omega/\sqrt{\epsinf}$ 
we have the following expression valid at low 
temperature ($k_B T \ll \hbar \omega_o$)
(for details see Appendix E)
\begin{equation} \label{eqn:gr_approx}
\tilde{g}(r) \simeq \frac{1}{\epsbar}
\left[
\frac{r}{\frac{\hbar}{m \omega_{LO}}} \left(
1 - \mbox{erf} \sqrt{\frac{r^2}{
\left< u^2 \right>
} }
\right)
+
\frac{2 r^2}{\left< u^2 \right>}
\frac{e^{- \frac{r^2}{2 \left< u^2 \right>} }}
{{ \left(2 \pi \left< u^2 \right> \right)}^{3/2}}
\right].
\end{equation}

The first term of eq. (\ref{eqn:gr_approx}) takes into account quantum charge
fluctuations which are  relevant at small distances, while the remaining term is
a classical contribution coming from the static charge distribution.
Notice that only the first term depends on the e-ph interaction.
Therefore, the polaron radius tends to the same high density asymptotic value
 for different values of the e-ph coupling $\alpha$
(see upper panel of fig.\ref{fig:radius_09}).

As a last point  we notice that the cross-over condition, roughly  estimated
as  $\omega_P \sim \omega_{LO} \sim \ (1-\eta)^2/\alpha^2 \sim 0.01/\alpha^2$, 
shifts toward higher densities as the e-ph coupling constant $\alpha$ is
reduced. In the weak coupling regime it lies in the liquid phase where RPA 
approaches in both 3D refs.\cite{Mahan,daCosta}  and in 2D case
\cite{Xiaoguang}  can be applied.   It is also worth to remark that for high
polarizability and for  all coupling the polaronic crossover is located in the 
liquid phase according to the highest value  of $\omega_{LO} \sim 
0.7/\alpha^2$ .

\section{III - 2D case} 

The results obtained in the 2D case are qualitatively similar to the 
3D case. Both the cross-over phenomenon in the low polarizability case 
and  the softening of the polaronic frequency in the 
high polarizability case are observed. 
Results are reported in the zero temperature phase diagram of fig.
\ref{fig:3D2D_summary}. In this figure, we compare the phase diagrams in 2D and 3D
by scaling appropriately the 2D e-ph coupling constant following
the single polaron results of
ref.\cite{scaling_Devreese}:
$\alpha_{3D} = (3 \pi / 4) \alpha_{2D}$. 
2D and 3D melting curves scale well according to 
the zero density scaling for all studied cases. A discrepancy is found in the  
the high polarizability strong e-ph coupling softening of $\omega_{pol}$.
Let us first discuss the scaling at finite density.
 
\begin{figure}[hbtp]
\begin{center}
\includegraphics[scale=.4,angle=0.]{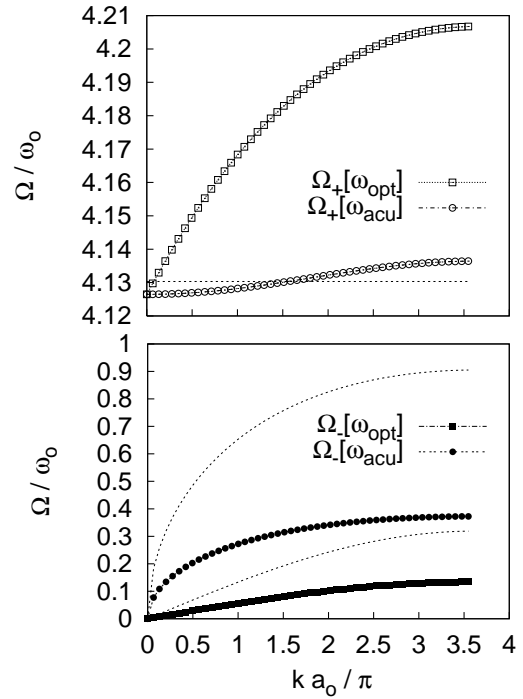}
\caption{
2D case. The eigenfrequencies of system along the direction (10) 
for $\eta=0.9$ and $\alpha=2.12$.
$1/r_s = 8 \; 10^{-3}$ (a.u.), $T=2.5 \; 10^{-5}$ (a.u.).
Density is close to the classical liquid-solid transition.
Upper panel: The frequencies of polaronic branch 
weakly dispersed around the polaronic frequency $\omega_{pol}\simeq v$ 
(dashed line). 
Lower panel: The renormalized Wigner crystal branches (points)
and the pure WC branches (dashed lines).}
\label{fig:2D_frequenze_09}
\end{center}
\end{figure}

In our variational scheme, the DOS of the WC is the peculiar difference between
the 2D and 3D cases. To see this explicitly let us compare the e-ph interaction
terms $\mathcal{S}^{self}_{e-ph-e}$.  Assuming polaronic units we get:
\begin{eqnarray}
\frac{1}{\beta}
\frac{ {\left< \mathcal{S}^{self}_{e-ph-e} \right>}_{T,3D} } {3 N} & = &  
- (\alpha) \frac{\sqrt{2}}{6}
\int_0^{\frac{\beta}{2}} \!\!\! d\tau
\frac{D_o(\tau)}{
\sqrt{ \frac{\pi}{2} d_{3D}(\tau) } } \label{eqn:scaling_I} \\
\frac{1}{\beta} \frac{ {\left< \mathcal{S}^{self}_{e-ph-e} \right>}_{T,2D} } {2 N}
& = &
-  
\left( \frac{3 \pi}{4} \alpha \right) 
\frac{\sqrt{2}}{6} 
\int_0^{\frac{\beta}{2}} \!\!\! d\tau
\frac{D_o(\tau)}{
\sqrt{ \frac{\pi}{2} d_{2D}(\tau) } } \nonumber \\
& & \label{eqn:scaling_II}
\end{eqnarray}
where the imaginary-time diffusion $d(\tau)$ eq.(\ref{eqn:d_tau}) is
itself a functional of the DOS. We notice from
eqs.(\ref{eqn:scaling_I},\ref{eqn:scaling_II}) that the free energy functional
scales explicitly as in the  single polaron case \cite{scaling_Devreese}
by scaling the coupling constant $\alpha$.
Related to the different 2D and 3D DOS we remark the different
behavior of the frequencies of
the normal modes.
Noticeably, the  ``optical'' branches go
to zero as $\sim \sqrt{k}$ at long wavelengths \cite{Crandall}. 
As in the 3D case, the frequencies of the LPC are splitted
in four branches (fig.\ref{fig:2D_frequenze_09})
$\Omega_{\pm} [\omega_{acu}(s,\vec{k})]$ 
and $\Omega_{\pm} [\omega_{opt}(s,\vec{k})]$
according to the same equation of 3D 
(see fig.\ref{fig:2D_frequenze_09}), where  
the 2D value for $\omega_W$ is given in appendix B 
eq. (\ref{eqn:w_Wig_2D}).

Let us discuss the deviation from the the scaling at strong coupling,    
which we see  from fig.\ref{fig:3D2D_summary} in the density of the softening  
of the polaronic frequency $\omega_{pol}$. Actually we observe that a steep fall 
of the variational parameter $v(r_s)$ occurs
as density increases  determining the softening of $\omega_{pol}$.
Peculiar features of the DOS enters in the variational determination of $v(r_s)$
as can achieved by the following argument. First of all we assume that 
$w$ is very close to the value $\omega_{LO}$ at strong coupling. 
Then we notice that 
as in 3D high polarizability case
the renormalized plasma frequency is much less than the phonon frequency and the
discussion which follows eq. (\ref{eqn:eps_s_k}) holds
for all densities  lower than
critical density of the softening. 
In this case the spectrum is composed by the low energy branches (renormalized
WC)  and the by polaronic branches weakly dispersed around $v$  
(see also fig.\ref{fig:2D_frequenze_09}).
Using this results at low temperatures 
$(\beta \rightarrow \infty)$, the condition for the extrema of 
$\mathcal{F}_{V}$ reads:
\begin{equation} \label{eqn:v_self_eq}
1 - \frac{1}{v^{3/2}} \sqrt{\frac{\omega^2_P}{w \varepsilon_o}} \mathcal{M}_2 
\; + \;  g(\alpha, \eta,r_s,v) = 0
\end{equation}
where the first and second terms are the derivative of $\mathcal{F}_T$ 
eq.(\ref{eqn:F_T}) and $g$ is the derivative of 
eqs.(\ref{eqn:mean_S_self_3D},\ref{eqn:mean_S_self_2D},\ref{eqn:mean_S_T}) 
from Appendix C. 
As $v(r_s) \rightarrow 0$  for $r_s \simeq r_c$ 
the second term acquires importance and DOS enters in the second moment 
$\mathcal{M}_2$. However there are other terms which are divergent as 
$v \rightarrow 0$ coming from the explicit form of the function $g$.

\section{Conclusions} 

We have studied the behavior of a low density electron gas in the presence of
a polarizable medium, where polaronic effects may play a relevant role. 
To determine the transition line, we have used a generalized Lindemann 
criterion which reproduces correctly the pure electron gas quantum melting. 
Because the amplitude of quantum fluctuations
depends on the e-ph renormalized plasma frequency, the
Lindemann rule  has been critically re-examined and adapted to  the polaron
crystal. This procedure allows to determine quantitatively the phase diagram of
the model and to extend the study of the model to the low polarizability case,
which was not studied before. 
We have also studied the 2D case, showing that the dimensional dependence 
is  not crucial to determine the 
nature of the quantum melting within in our harmonic variational  scheme.
The scaling predicted for the e-ph coupling constant at zero density do 
apply as well at non zero density up to quantum melting.
A noticeable difference instead is in the position of the 
density of the softening of the polaronic frequency 
which in 3D is much closer to the  melting 
than in 2D case.  This suggests that the hetero-interactions  are less
effective to destabilize the dipolar crystal in 2D. However other possible
mechanism, (lattice-effects,structural disorder  or impurities) can cooperate
to the localization together with  the interactions between the electrons and
lead to the formation  of a pinned Wigner Crystal.   In this case the melting
can not be predicted by the Lindemann rule but a similar dipolar instability
due to the long-range interaction  between the electrons can still drives the
melting.

While the weak e-ph coupling regime is similar for both low and high
polarization case, the strong coupling scenario is qualitatively different. 

In the high polarizability regime, we have recovered the incipient instability
which was found in previous studies near the solid phase 
\cite{Simone_meanfield,Simone} and also in the liquid phase \cite{Lorenzana}.
In comparison with previous work, we have found that this regime is restricted
to very large values of the coupling $\alpha>10$ leaving an interesting
intermediate region of coupling in which polarons may exist in the liquid
phase. This region can in principle be explored with non perturbative 
numerical techniques as e.g. Path Integral Monte Carlo. Work along this
direction is currently in progress.

In the low polarizability regime, a crossover occurs inside the solid phase when 
the renormalized plasma frequency approaches the phonon frequency.  
At low density, we still have a LPC, while at higher
densities the electron-phonon interaction is weakened irrespective of the {\it
bare} electron-phonon coupling. In this case, polaron clouds overlap as
and the polaron feature of the crystal is lost. 
The crossover from polaronic $(\omega_P < \omega_{LO})$ to 
non polaronic $(\omega_P > \omega_{LO})$ character has been 
observed in weakly coupled systems  
such as GaAs in the liquid phase and analyzed in term RPA 
\cite{Mahan,Xiaoguang}. In this system it occurs around 
$r_s \sim 0.6-0.7$ while for ZnO $\alpha$ is larger shifting 
the crossover to $r_s \sim 7$.
Finding a system with low polarizability and larger e-ph 
coupling is difficult since it implies very 
low $\omega_{LO}$: $\alpha \sim  (1 - \eta) / \omega_{LO}$  
from eq.(\ref{eqn:alpha}).
However in surfaces of InSb an $\alpha = 4.5$ has been 
predicted together with $\eta= 0.88$ \cite{Sun_Gu} 
leading to the possible observation of the 
crossover inside the solid phase.

We notice also that our low polarizability scenario of density crossover
inside the solid phase bear some resemblance to that found for ripplonic
polaron systems \cite{Andrei_book}. 
Though the electron-ripplon interaction in these systems is different from the 
Fr\"ohlich type, resonances in the absorption spectrum observed by Grimes and
Adams \cite{Grimes_Adams}, their explanation at low density 
\cite{Fisher_Halperin} relays on
the same qualitative arguments developed in the present work.
Recent works on high density ripplonic polaron systems realized on a helium
bubbles predicts also in this case a mixing between plasmon and polaron modes
 \cite{DeVreese-multiripplons}.

Finally we remark that we have obtained an appreciable stabilization of the 
crystal phase even for intermediate regime $\alpha \sim 3-5$ in low  
polarizability cases. 
We conclude that the general result that e-ph interaction effects can stabilize the Wigner
crystal phase could motivate experimental studies on two dimensional electronic
devices involving polarizable media.
To this aim a layered configuration is advised even with some 
warnings  \cite{Fomin_Smondyrev}. In 2D heterostructure the use 
of a perpendicular electric field  \cite{Fomin_Smondyrev} 
could not only increase the 
polaron effect but also tune it as it was shown in the 
case of charged helium surfaces \cite{Helium_surface}.

\section{ACKNOWLEDGMENTS}

Authors acknowledge S. Fratini and P. Quemerais for useful discussions and 
critical reading of the manuscript. We thanks also J. Lorenzana for useful
suggestions.
One of us (G.R.) thanks the also the kind 
hospitality of CNRS-LEPES Grenoble were a part of this work has been done.

This work was supported by MIUR-Cofin 2001 and MIUR-Cofin 2003 matching funds
programs.

\section{APPENDIX A: The low energy cut off in 2D} 

The 2d DOS function can be defined as

\begin{equation}
\rho(\omega)  =  \sum_{s=1,2} \int_{V_ {B}} \!\!\!\! d^2 k \; \delta (\omega - \omega_{s,k})
\end{equation}

where  $V_ {B}$ is the volume of the first Brillouin`s zone $(1BZ)$. 
Let us consider a small fraction $\varepsilon$ of the plasma frequency $\omega_P$.
At long wavelength $(k=0)$, we have the 2D dispersion laws for the acoustical mode is
$\omega_1 (k)  \simeq  c_1 k $ while the "optical" $\omega_2 (k) \simeq  c_2 \sqrt{k}$ \cite{Bonsall}.
As a consequence the behavior of the DOS for $\omega \simeq 0$ is

\begin{eqnarray}
\int^{2 \pi}_0 \!\!\!\! d \theta \int^{k(\varepsilon)}_0 \!\!\!\!\!\!\!  
d k \; k \; \delta (\omega - c_1 k) &=&
\frac{2 \pi}{c^2_1} \omega \\
\int^{2 \pi}_0 \!\!\!\!   d \theta \int^{k(\varepsilon)}_0 \!\!\!\!\!\!\!   d k \; k \;
\delta (\omega - c_2 \sqrt{k}) &=&
\frac{4 \pi}{c^4_2} \omega^3 
\end{eqnarray}

Introducing the scaled frequency $x$ defined as  $\omega = \omega_P x$ and  
the quantum parameter $\eta_q$ eq.(\ref{eqn:eta_q}), the thermal electronic fluctuation 
is expressed as an average on the DOS as
\begin{equation}  \label{eqn:u2DOS}
\left< u^2 \right>  
=  \frac{\hbar}{m} {\left< \frac{\coth (\eta_q x)}{x} \right>}_{DOS} 
\end{equation} 
\begin{equation} \label{eqn:u2_expansion} 
 =  \frac{\hbar}{m} \int \!\!\! d  x \; \rho(\omega_P x) 
\left[ \frac{1}{\eta_q x^2} + 
\eta_q  \left( \alpha_0 + \alpha_2 {\left( \eta_q x \right)}^{2} + \dots \right) 
\right] 
\end{equation} 
Since $\rho(\omega_P x) \sim x$ for $x \rightarrow 0$,
the average in eqs. (\ref{eqn:u2DOS})
converges for any of $x^{2n}$ with $n \ge 0$
in the the expansion eq.(\ref{eqn:u2_expansion}).
In the $n=-1$ term we consider infrared cut-off $x_c$
giving

\begin{eqnarray}
{\left< \frac{1}{x^2} \right>}_{DOS}  &\simeq&
\int^{\varepsilon}_{x_c} \!\!\!\!\! d x \; \frac{(2 \pi / c^2_1) \omega_P}{x}  +
\int_{\varepsilon} \!\!\! d x \; \frac{\rho(\omega_P x)}{x^2} \nonumber \\
&\simeq &
\frac{2 \pi \omega_P}{c^2_1} \ln \left( \frac{\varepsilon}{x_c} \right)
+
\int_{\varepsilon} \!\!\!  d x \; \frac{\rho(\omega_P x)}{x^2}
\end{eqnarray}

This term diverges logarithmically as $x_c \rightarrow 0$.
However $\eta_q \rightarrow \infty$ as we approach  the quantum region. 
The electronic fluctuation turns out to be cut-off independent if

\begin{equation}
\left< \frac{1}{x^2} \right> \ll
\eta^2_q  \left<
\alpha_0 + \alpha_2 {\left( \eta_q x \right)}^{2} + \dots
\right>. \label{eqn:mom_inv2_vs_quantum_corr}
\end{equation}

We have chosen for  the cut-off frequency 
$x_c = \omega_{min} / \omega_P \simeq 5 \cdot  10^{-5}$
so that  the condition
eq.(\ref{eqn:mom_inv2_vs_quantum_corr}) is fulfilled
around $\eta_q(T,r_s) \ge 10$ which  corresponds to
a large region inside to the solid phase. 
By the relation for acoustical long wave excitation
$\omega_{min} = c_1  k_{min}$ and
$k_{min} =   2 \pi / \left( r_s \sqrt{N} \right)$,
the number of electrons is $N=  5.24 \cdot 10^{6}$.
Our inverse second moment of DOS is $\mathcal{M}_{-2} = 12.5$ 
(cfr ref. \cite{Gann} $\mathcal{M}_{-2}=8.16$ for $N=1024$).

\section{APPENDIX B: The harmonic variational approximation}
We expand in the harmonic approximation the terms 
$\mathcal{S}_{e-e},\mathcal{S}_{e-J},\mathcal{S}^{dist}_{e-ph-e}$
(eqs.\ref{eqn:S_dist_e-e},\ref{eqn:S_e-J},\ref{eqn:S_dist_e-ph-e}). 
Let $\vec{r}_{\imath} = \vec{R}_{\imath} + \vec{u}_{\imath}$, where
$\vec{R}_{\imath}$ is the lattice point of the crystal and
$\vec{u}_{\imath}$ is  the electronic displacement from $\vec{R}_{\imath}$,
and set $\Delta \vec{u}_{\imath,\jmath}(\tau,\sigma) = \vec{u}_{\jmath}(\sigma) -\vec{u}_{\imath}(\tau)$
and  $\vec{R}_{\jmath,\imath} = \vec{R}_{\jmath} - \vec{R}_{\imath}$.
The static terms give

\begin{equation}
\mathcal{S}^o_{e-e}(\{ \vec{R}_{\imath} \})
+ \mathcal{S}^o_{e-J}(\{ \vec{R}_{\imath} \}) 
+ \mathcal{S}^{o,dist}_{e-ph-e}(\{ \vec{R}_{\imath} \}) =
\frac{\mathcal{S}^{o}_{WC}(\{ \vec{R}_{\imath} \})}{\epszero}
\end{equation}

the e-ph interaction does not change the equilibrium positions  of the pure
electronic crystal (WC) which corresponds to the minimum of the potential
energy.\\
The sum of the dynamical parts in the harmonic approximation gives

\begin{eqnarray}
\mathcal{S}^{H}_{e-e} + \mathcal{S}^{H}_{e-J} + \mathcal{S}^{H,dist}_{e-ph-e} = \nonumber \\
 = \int^{\beta}_{0} \!\!\!\! d \tau  \sum_{\imath}
\left[ V_{J} (\vec{u}_{\imath}(\tau)) +  V_{e-e} (\vec{u}_{\imath}(\tau))
\right] \label{eqn:S_dyn}
\end{eqnarray}
where
\begin{eqnarray}
V_{J} (\vec{u}_{\imath}(\tau))  & = &
- \frac{e^2 \rho_J}{\epszero} \int \!\!\! d^D \! r 
\left( \frac{1}{|\vec{u}_{\imath}(\tau) - \vec{r}|} - \frac{1}{r} \right) \label{eqn:V_J}\\
V_{e-e} (\vec{u}_{\imath}(\tau)) &=&
\frac{e^2}{2}
\int^{\beta}_{0} \!\!\!\!  d \sigma F(\tau -\sigma)  \sum_{\jmath \neq \imath}
U_{\jmath,\imath}(\tau,\sigma)  \label{eqn:V_H_ij} \\
U_{\jmath,\imath}(\tau,\sigma) &=&
 \frac{1}{2}
\Delta \vec{u}_{\imath,\jmath}(\tau,\sigma)
\cdot \overline{\mathcal{I}}(\vec{R}_{\jmath,\imath})
\Delta \vec{u}_{\imath,\jmath}(\tau,\sigma)  \label{eqn:U_H_ij} \\
F(\tau -\sigma) &=&
\frac{\delta(\tau-\sigma)}{\epsinf} - \frac{\omega_{LO}}{2 \epsbar} D_o(\tau - \sigma) \\
{\left[ \overline{\mathcal{I}_{\imath \jmath}} \right]}_{\alpha \beta} &=&
\frac{\delta_{\alpha \beta}}{{\left| \vec{R}_{\jmath,\imath} \right|}^3} -
3 \frac{{\left[ \vec{R}_{\jmath,\imath} \right]}_{\alpha}
{\left[\vec{R}_{\jmath,\imath}\right]}_{\beta}}
{{\left| \vec{R}_{\jmath,\imath} \right|}^5} \label{eqn:I_ij}
\end{eqnarray}

>From now we drop on the double $\sigma,\tau$ indexes in $\Delta \vec{u}_{\imath,\jmath}$.
To evaluate the integral in eq.(\ref{eqn:V_J}) and the sums on index $\jmath$  in eq.(\ref{eqn:V_H_ij}),
we consider  a sphere $S_R$ of radius $R_s$ (a disk in 2D) centered on site $\imath$. 
We first sum on index $\jmath$ and then we perform the limit $R_s \rightarrow \infty$.
Finally we sum on index $\imath$ in eq. (\ref{eqn:S_dyn}).

\subsection{3D case}

By Gauss's law (with the condition $V_{J}(0)=0$), we have 
\begin{equation} \label{eqn:Wigner_3D}
V_{J} (\vec{u}_{\imath}(\tau))  = 
\frac{1}{2} m \frac{\omega^2_W}{\epszero} {\left|\vec{u}_{\imath}(\tau)\right|}^2
\end{equation}
where the Wigner frequency is $\omega^2_W = \omega^2_P / 3$.
Because of $V_{J}$ is independent of the size
of $S_R$, eq.(\ref{eqn:Wigner_3D})  does not change in the
limit $R_s \rightarrow \infty$.\\
To evaluate the sum in eq.\ref{eqn:V_H_ij} with the condition $R_{\jmath} < R_s$
we remind that we have two self terms ($(i,i)$ and $(j,j)$) and two distinct terms
($(i,j)$ and $(j,i)$) in eq. (\ref{eqn:U_H_ij}).
The two self terms give the same contribution ,as can be easily check
if we firstly we carry on the limit $R_s \rightarrow \infty$ and
then the sum on index $\jmath$ and $\imath$. They vanishes
because of cubic symmetry of the lattice. 
When the two distinct terms ($(i,j)$ and $(j,i)$) of eq. (\ref{eqn:U_H_ij}) 
are  inserted in eq.(\ref{eqn:V_H_ij}) and the limit $R_s \rightarrow \infty$ 
is taken, we obtain the term $V_{e-e}(u_{\imath})$ of eq.\ref{eqn:S_dyn}

\begin{equation} \label{eqn:U_H_ij_R_sum_III}
V_{e-e}  (\vec{u}_{\imath}(\tau)) =
\frac{e^2}{2} \sum_{\jmath \neq \imath}
\int^{\beta}_0 \!\!\!\! d \sigma
F(\tau - \sigma)
\vec{u}_{\jmath}(\sigma)
\overline{\mathcal{I}}(\vec{R}_{\jmath,\imath})
\vec{u}_{\imath}(\tau)   
\end{equation}

Summing on index $\imath$ and integrating on variable $\tau$ 
the eqs.(\ref{eqn:Wigner_3D},\ref{eqn:U_H_ij_R_sum_III}),
we obtain the terms $\mathcal{S}^H_{e-J} , \mathcal{S}^{H}_{e-e}, \mathcal{S}^{H,dist}_{e-ph-e}$
eqs.(\ref{eqn:S_H_dist_e-e},\ref{eqn:S_H_dist_e-ph-e}).

\subsection{2D case}

In 2D the interaction potential  $V^{R}_{J}(u)$
of  a uniform positive charged disk of radius
$R_s$ eq. (\ref{eqn:V_J}) is

\begin{equation}
V^{R}_{J}(u) = 
- \frac{e^2 \rho_J}{\epszero} \int^{2 \pi}_0 \!\!\!\!\!\! d \theta F(\theta) \\
\end{equation}
where
\begin{eqnarray*}
& F(\theta) =  \sqrt{R^2_s + u^2 -2 R_s u \cos(\theta)} - u - R_s &\\
&+ u \cos(\theta) \ln \frac{R_s - u \cos(\theta) + \sqrt{R^2_s + u^2 -2 R_s u 
\cos(\theta)}}{u(1-\cos(\theta))}&
\end{eqnarray*}

In the limit $R_s \rightarrow \infty$
\begin{equation}
\lim_{R \rightarrow \infty} V^{R}_{J}(\vec{u}_\imath)  =
\lim_{ \frac{u}{R} \rightarrow  0}
\frac{e^2}{\epszero} \rho_J \frac{\pi}{R_S} u^2_{\imath} = 0
\end{equation}
since the total electric field of an infinite charged disk
is perpendicular to the disk.\\
Then we have to evaluate the sums eq.(\ref{eqn:U_H_ij}).
The two distinct terms ($(i,j)$ and $(j,i)$) gives the identical 
result eq.\ref{eqn:U_H_ij_R_sum_III} of the 3D case while
the self term $(\imath,\imath)$  is written as

\begin{equation}
\frac{1}{2} \vec{u}_{\imath} \left( 
\sum_{ \stackrel{\vec{R}_{\jmath} < R}{\jmath \neq \imath} }
\overline{\mathcal{I}_{\imath \jmath}}
\right) \vec{u}_{\imath}  = \vec{u}_{\imath} \overline{\mathcal{D}} \vec{u}_{\imath}
\end{equation}

the matrix $\mathcal{D}$ in 2D is
defined as sum of the matrices $\overline{\mathcal{I}}(\vec{R}_\jmath)$
eq.(\ref{eqn:I_ij}) on hexagonal lattice points $\vec{R}_\jmath$. 
Contrary to the 3D case, the matrix $\mathcal{D}$ is not zero in 2D case.
By the lattice symmetry, the off-diagonal elements are zero while 
the diagonal terms are equal to the local potential which acts on each electrons 

\begin{equation} \label{eqn:Wigner_2D}
\frac{e^2}{2}\int^{\beta}_{0} d \sigma F(\tau - \sigma)
\overline{\mathcal{D}}_{\alpha \alpha} {\left|\vec{u}_{\imath}(\tau)\right|}^2  =
\frac{1}{2} m \frac{\omega^2_W}{\epszero} {\left|\vec{u}_{\imath}(\tau)\right|}^2 
\end{equation}
where we use as definition 2D Wigner frequency
\begin{equation} \label{eqn:w_Wig_2D}
\omega^2_{W} = \frac{e^2}{m}
\lim_{R \rightarrow \infty}
\sum_{\stackrel{\jmath \neq \imath}{R_{\jmath,\imath}< R}}
\frac{1}{2 R^3_{\jmath,\imath}} \qquad \mbox{(2D)}
\end{equation}

For an hexagonal lattice of nearest neighbor distance $d_{n.n.}$,
we have $\sum_{\jmath \neq \imath}(1/2 R^3_{\jmath,\imath}) =  5.51709 / d^3_{n.n.}$.\\
Summing on index $\imath$ eqs.(\ref{eqn:U_H_ij_R_sum_III},\ref{eqn:Wigner_2D})
we obtain the terms $\mathcal{S}^H_{e-J} , \mathcal{S}^{H}_{e-e}, \mathcal{S}^{H,dist}_{e-ph-e}$
eqs.(\ref{eqn:S_H_dist_e-e},\ref{eqn:S_H_dist_e-ph-e}).

\subsection{Normal modes}

The WC normal modes are defined as
\begin{equation} \label{eqn:q_ks}
\vec{u}_{\imath} = 
\frac{1}{\sqrt{N}} \sum_{\vec{k},s}
\hat{\varepsilon}_{\vec{k},s }
q_{\vec{k},s}
e^{\imag \vec{k} \vec{R}_{\imath}}
\end{equation}
where the vectors $\vec{k}$ belongs to the $1BZ$ of the reciprocal lattice, 
$\hat{\varepsilon}_{\vec{k},s}$ are
eigenvector with eigenvalue $\omega^2_{\vec{k},s}$
of the dynamical matrix $\overline{\mathcal{M}}$
which is defined as
\begin{equation} \label{eqn:M_dyn}
{\overline{\mathcal{M}}}_{\alpha \beta}   = \delta_{\alpha \beta}
\omega^2_W  + \frac{e^2}{m}
\sum_{\vec{R}_{\imath} \neq 0}
{\overline{\mathcal{I}}}_{\alpha \beta} (\vec{R}_{\imath})
 e^{\imag \vec{k} \cdot \vec{R}_{\imath}}
\end{equation}
Inserting the WC normal modes eq.(\ref{eqn:q_ks})
in eqs.(\ref{eqn:S_Fey},\ref{eq:SK},\ref{eqn:S_H_dist_e-e},\ref{eqn:S_H_dist_e-ph-e}),
we express the harmonic variational action $\mathcal{S}_T$ as
\begin{equation} \label{eqn:S_T_q}
\mathcal{S}_{T} (\{ q_{s,\vec{k}}(\tau) \})=
\sum_{s,\vec{k}} \int_0^{\beta} \!\!\! d \tau L_{s,k} \left( \tau \right)
\end{equation}
where the Lagrangian is
\begin{equation} \label{eqn:L_sk}
\begin{array}{ll}
L_{s,k}   &=
\frac{1}{2} m {\left| \dot{q}_{\vec{k},s} (\tau) \right|}^{2} +
\frac{1}{2} m \frac{\omega^2_{\vec{k},s}}{\epszero}
{\left|q_{\vec{k},s} (\tau) \right|}^2  \\
&+
 \frac{m w  \left(v^2 - w^2\right)}{8}
\displaystyle{\int_0^{\beta} }   \!\!\!\! d\sigma 
D_V(\tau-\sigma) 
{\left| q_{\vec{k},s} (\tau) -  q_{\vec{k},s} (\sigma) \right| }^2  \\
&+
\frac{m \omega_{LO} \left(\omega^2_{\vec{k},s} -\omega^2_W \right)}{8 \epsbar}
\displaystyle{\int_0^{\beta} }  \!\!\!\! d\sigma D_o(\tau-\sigma) 
{\left|q_{\vec{k},s} (\tau)- q_{\vec{k},s} (\sigma) \right|}^2    \\
\end{array} 
\end{equation}

\section{APPENDIX C: The variational free energy $\mathcal{F}_V$}

The first term of the variational free energy  $\mathcal{F}_V$
eq. (\ref{eqn:F_harm_var})  is the free energy  $\mathcal{F}_T$
associated to the partition function of the trial action  $\mathcal{Z}_T$.
This is calculated as the functional integral eq.(\ref{eqn:Z_eff}) 
where  $\mathcal{S}_{eff}$ eq. (\ref{eqn:S_eff}) is replaced 
by $\mathcal{S}_T$ eq. (\ref{eq:Svar}).
The second term of $\mathcal{F}_V$  is the mean value eq. (\ref{eqn:mean})
of the difference  between $\mathcal{S}^{self}_{e-ph-e}$
eq. (\ref{eqn:S_self_e-ph-e}) and  $\mathcal{S}_{Feyn}$
eq. (\ref{eqn:S_Fey}).\\
We start by changing the dynamical variables of integration from
$\{ \vec{u}_\imath (\tau) \}$  to $\{ q_{s,\vec{k}} (\tau)  \}$.
By reality condition we have $q_{-\vec{k},s} = q^{\ast}_{\vec{k},s}$
and $\hat{\varepsilon}_{-\vec{k},s }=- \hat{\varepsilon}_{\vec{k},s}$,
we must sum only  $\vec{k}$ vectors in the upper half space ($k_z > 0$) of $1BZ$
\begin{equation} \label{eqn:q_ks_II}
\vec{u}_{\imath} =
\frac{1}{\sqrt{N}}\sum_{s,\vec{k},k_z > 0}
\hat{\varepsilon}_{\vec{k},s }
\left[ q_{\vec{k},s} e^{\imag \vec{k} \vec{R}_{\imath}} -
 q^{\ast}_{\vec{k},s} e^{- \imag \vec{k} \
\vec{R}_{\imath}} \right]
\end{equation}
Therefore the real and imaginary part of $q_{s,\vec{k}}$ for all $k$ with $(k_z >0)$
of the $1BZ$ are the actual independent variables and the Jacobian 
of canonical transformation is $J=2^{DN}$

\begin{equation} \label{eqn:Z_T}
\mathcal{Z}_T = \int \!\! J \!\! \prod_{s,\vec{k},k_z > 0}
\mathcal{D} [q^{Re}_{s,\vec{k}} (\tau)]  \mathcal{D} [q^{Im}_{s,\vec{k}} (\tau)]
e^{- \mathcal{S}_{T} \left[ \{ q_{s,\vec{k}}(\tau) \} \right] }
\end{equation}

Using the periodicity condition $(q_{s,\vec{k}}(0)~=~q_{s,\vec{k}} (\beta))$,
we have the following Fourier expansion $(\omega_n = (2 \pi/\beta) n)$
\begin{eqnarray}
q_{s,\vec{k},n} &=&
\frac{1}{\beta} \int^{\beta}_0 \!\! d \tau \; q_{s,\vec{k}}(\tau) e^{- \imag \omega_n \tau} \nonumber \\
q_{s,\vec{k}}(\tau)        & = &  q_{s,\vec{k},c} + \delta q_{s,\vec{k}}(\tau)  \\
q_{s,\vec{k},c}            & = &
\frac{1}{\beta} \int^{\beta}_0  d \tau \; q_{s,\vec{k}}(\tau) \label{eqn:centroid}\\
\delta q_{s,\vec{k}}(\tau)
& = & \sum^{\infty}_{\stackrel{n= - \infty}{n \neq 0}} q_{s,\vec{k},n}
e^{\imag \omega_n \tau} \label{eqn:fluctua}
\end{eqnarray}
where we have separated the mean value 
of path on the imaginary time eq.(\ref{eqn:centroid}) (centroid)
from the fluctuation around it eq.(\ref{eqn:fluctua}).
The action $\mathcal{S}_{T}(\{ q_{s,\vec{k}}(\tau) \})$ is quadratic in
$\{ q_{s,\vec{k},n} \}$  therefore we can separate eq.(\ref{eqn:Z_T})
in two gaussian integrals
\begin{equation}
\mathcal{Z}_T = \mathcal{Z}_{T,c}  \mathcal{Z}_{T, \delta q}
\end{equation}
\begin{eqnarray}
\mathcal{Z}_{T,c}  &=&   \!\! \int \!\!   \prod_{s,\vec{k},k_z > 0}
\frac{d q^{Re}_{\vec{k},s,c} d q^{Im}_{\vec{k},s,c}}{\pi \hbar^2 / m k_B T}
\;\; e^{- \mathcal{S}^{c}_{T} \{ q_{s,\vec{k},c} \}}  \nonumber \\
& = &  \int  \prod_{s,\vec{k},k_z > 0}
\frac{d q^{Re}_{\vec{k},s,c} d q^{Im}_{\vec{k},s,c}}{\pi \hbar^2 / m k_B T}
\;\;
e^{-\frac{m}{k_B T}  \!\!\!
\frac{{\left| q_{\vec{k},s,c} \right|}^2}{\omega^2_{s,\vec{k}}/ \epszero}
} \nonumber \\
&=&
\prod_{s,\vec{k}} \frac{k_B T} {\hbar \; \omega_{s,\vec{k}} / \sqrt{\epszero} } \label{eqn:Z_T_C}
\end{eqnarray}
hence after we omit the classic term  $\mathcal{Z}_{T,c}$ eq.(\ref{eqn:Z_T_C}).
\begin{eqnarray}
\mathcal{Z}_{T, \delta q}  &=&    \!\! \int \!\!
  \prod_{\stackrel{n \neq  0}{s,\vec{k},k_z > 0}}
\frac{d q^{Re}_{\vec{k},s,n}  d q^{Im}_{\vec{k},s,n} }{\pi k_B T / m \omega^2_n}  \;\;
e^{- \delta \mathcal{S}_{T}  \{ \delta q_{s,\vec{k}}(\tau)  \}  } \nonumber \\
&=&
\int  \prod_{\stackrel{n \neq  0}{s,\vec{k},k_z > 0}}
\frac{d q^{Re}_{\vec{k},s,n}  d q^{Im}_{\vec{k},s,n} }{\pi k_B T / m \omega^2_n}
\;\; e^{-\frac{m}{k_B T}
\frac{{\left| q_{\vec{k},s,n} \right|}^2}{\lambda_{s,\vec{k},n}}
}  \nonumber  \\
&=& \prod_{\stackrel{n \neq  0}{s,\vec{k},k_z > 0}}
\omega^2_n \lambda_{s,\vec{k},n}  \label{eqn:Z_del_A}
\end{eqnarray}
where
\begin{eqnarray}
\lambda_{s,\vec{k},0} &=& \frac{1}{\omega^2_{\vec{k},s} / \epszero} \\
\lambda_{s,\vec{k},n} & = & 
\sum^{3}_{\gamma=1} \frac{A_{\gamma}}{\omega^2_n +\Omega^2_{\gamma}} \\
A_{1} &=&
\frac{(\Omega^2_1 - \omega^2_{LO}) (\Omega^2_1 - w^2_T)}
{(\Omega^2_1 - \Omega^2_2)(\Omega^2_1-\Omega^2_3)} \quad (\mbox{cyclic perm.} \; \gamma=1,2,3) \nonumber \\
& & \label{eqn:A_factors} 
\end{eqnarray}
the  frequencies $\Omega^2_{\gamma} \; (\gamma=1,2,3)$  are  
the opposite of the roots of cubic
\begin{eqnarray}
\mathcal{P}_3 (z) & = &  z^3 + a_2 z^2 + a_1 z + a_0   \label{eqn:pol3}\\
a_2 & = & v^2 + \omega^2_{LO} + \frac{\omega^2_{\vec{k},s}}{\epszero} +
 \frac{\omega^2_{\vec{k},s} - \omega^2_W}{\epsbar}  \nonumber \\
a_1 & = & \omega^2_{LO} v^2 + \frac{\omega^2_{\vec{k},s}}{\epszero} (\omega^2_{LO} + w^2 ) +
 w^2 \frac{\omega^2_{\vec{k},s} - \omega^2_W}{\epsbar}  \nonumber \\
a_0 & = &  \frac{\omega^2_{LO} w^2 \omega^2_{\vec{k},s}}{\epszero}  \nonumber
\end{eqnarray}
The gaussian integrals  eq.(\ref{eqn:Z_del_A}) are convergent
if $\lambda_{s,\vec{k},n}$ are positive numbers
$\forall (s,\vec{k},n)$. This condition is fulfilled
if  $\Omega^2_{\gamma}$
are {\sl all} positive.
The numerical minimization of the variational free energy
has been made enforcing this constraint.
Performing the infinite product in eq.(\ref{eqn:Z_del_A}) we have 
\begin{eqnarray}
\mathcal{Z}_{T, \delta q} & = &
{\left( \frac{\sinh(\hbar \omega_{LO} /2 k_B T) }{\hbar \omega_{LO} /2 k_B T } \right)}^{DN}
{\left( \frac{\sinh(\hbar w /2 k_B T) }{\hbar w /2 k_B T } \right)}^{DN} \nonumber \\
&\cdot& \prod_{s,\vec{k},\gamma}
\frac{\hbar \Omega_{\gamma,s,\vec{k}} /2 k_B T }
{\sinh(\hbar \Omega_{\gamma,s,\vec{k}} /2 k_B T) } \nonumber
\end{eqnarray}
and finally we substitute the sum on $(\vec{k}_{\imath},s)$ 
with the integral on  the WC DOS $\rho(\omega)$ in the
free energy  $\mathcal{F}_T$
\begin{eqnarray} \label{eqn:F_T}
\frac{\mathcal{F}_T}{D N}
&=& - k_B T \ln \left[\sinh\left(  \frac{\hbar \omega_{LO}}{2 k_B T}  \right)
\sinh\left(  \frac{\hbar w}{2 k_B T} \right) \right] \nonumber  \\
& + & k_B T \int d \omega \rho(\omega)
\sum^{3}_{\gamma=1} \ln \left[ \sinh \left(
 \frac{\hbar \Omega_{\gamma} (\omega)}{2 k_B T}  \right) \right]
\end{eqnarray}

To calculate the mean value of $\mathcal{S}^{self}_{e-ph-e}$
eq. (\ref{eqn:S_self_e-ph-e}) in $3D$ we use the following identity \cite{footnote_D_o}
\begin{eqnarray}
\iint_0^{\beta} \!\!\! d\tau  d\sigma
 D_o(\tau-\sigma)
\int \!\! \frac{d^3 q}{ {\left( 2 \pi \right)}^3 } \frac{4 \pi}{q^2} 
{\left<
e^{\imag \vec{q} \cdot \left[ \vec{u}_{\imath}(\tau)-\vec{u}_{\imath}(\sigma) \right] }
\right>}_T &=&  \nonumber  \\
 =   - 2 \beta 
\int_0^{\frac{\beta}{2}} \!\!\! d\tau
\frac{D_o(\tau)}{
\sqrt{ \frac{\pi}{2} d_{3D}(\tau) } } & & \label{eqn:mean_S_self_3D}
\end{eqnarray}
while in $2D$ $(q^2 = q^2_{\perp} + q^2_z)$
\begin{eqnarray}
\iint_0^{\beta} \!\!\! d\tau d\sigma
D_o(\tau-\sigma)
\int \frac{d^2 q_{\perp}}{ {\left( 2 \pi \right)}^2 } 
 \frac{2 \pi}{q_{\perp}}
e^{- \frac{1}{2} d_{2D}(\tau - \sigma) q^2_{\perp}} &=&
\nonumber  \\
= - 2 \beta  \left( \frac{\pi}{2} \right)
\int_0^{\frac{\beta}{2}} \!\!\! d\tau
\frac{D_o(\tau)}{
\sqrt{ \frac{\pi}{2} d_{2D}(\tau) } } & & \label{eqn:mean_S_self_2D}
\end{eqnarray}
where $d_D (\tau)$ is the imaginary time diffusion in the LPC 
defined as (3D or 2D)
\begin{equation} \label{eqn:d_tau}
d_D (\tau) =\frac{ 
{\left<  {\left| \vec{u}(\tau) - \vec{u}(0)  \right|}^2 \right>}_T }
{D}
\end{equation} 
The mean value of  $\mathcal{S}_{Feyn}$
eq. (\ref{eqn:S_Fey}) is
\begin{eqnarray}
& &   {\left< \mathcal{S}_{Feyn} \right>}_T / N =   \nonumber \\
&=&  - D 
\frac{m w  \left(v^2 - w^2\right)}{8}
\iint_0^{\beta} \!\! d\tau  d\sigma  
D_T(\tau-\sigma)
d_D (\tau - \sigma) \nonumber \\
& & \label{eqn:mean_S_T}
\end{eqnarray}
To obtain eqs.(\ref{eqn:mean_S_self_3D},\ref{eqn:mean_S_self_2D},\ref{eqn:mean_S_T})
we have used
\begin{equation} \label{eqn:Y_tau}
{\left<
e^{\imag \vec{q} \cdot \left[ \vec{u}_{\imath}(\tau)-\vec{u}_{\imath}(\sigma) \right] }
\right>}_{T}   = e^{- \frac{1}{2} d_D (\tau - \sigma) q^2 }
\end{equation}
We will demonstrate eq.(\ref{eqn:Y_tau}) in the next subsection.

\subsection{Calculation of
${\left<\exp(\imag \vec{q} \cdot \left[ \vec{u}_{\imath}(\tau)-\vec{u}_{\imath}(\sigma) \right])
\right>}_T$ }

>From eqs.(\ref{eqn:q_ks},\ref{eqn:fluctua}) we have
\begin{equation} \label{eqn:mean_exp_I}
\imag \vec{q} \cdot \left[ \vec{u}_{\imath}(\tau)-\vec{u}_{\imath}(\sigma)  \right]
 = \sum_{\stackrel{s,k_z>0}{n \neq 0}}
\left[
 q_{\vec{k},s,n} J^{\ast}_{s,k,n}(\tau-\sigma,\vec{q})
+
\mbox{c.c.}
\right]
\end{equation}
\begin{equation} \label{eqn:mean_exp_II}
J^{\ast}_{s,k,n}(\tau-\sigma,\vec{q}) =
\frac{\imag}{\sqrt{N}}
\vec{q} \cdot  \hat{\varepsilon}_{\vec{k},s }
\left( e^{\imag \omega_n \tau} - e^{\imag \omega_n \sigma} \right)
e^{\imag \vec{k} \vec{R}_{\imath}}
\end{equation}
then we have
\begin{eqnarray}
& & {\left<\exp(\imag \vec{q} \cdot \left[ \vec{u}_{\imath}(\tau)-\vec{u}_{\imath}(\sigma) \right])
\right>}_T = \nonumber \\
&=&
\frac{1}{\mathcal{Z}_{T, \delta q}} \!\!\! \int \!\!\!\!\!\! 
\prod_{\stackrel{n \neq  0}{s,\vec{k},k_z > 0}} \!\!\!
\frac{d q^{Re}_{\vec{k},s,n}  d q^{Im}_{\vec{k},s,n} }{\pi k_B T / m \omega^2_n}  \;\;
e^{-\frac{m}{k_B T}
\frac{{\left| q_{\vec{k},s,n} \right|}^2}{\lambda_{s,\vec{k},n}}
+
q_{\vec{k},s,n} J^{\ast}_{\vec{k},s,n} + \mbox{c.c.} } \nonumber \\
&=& \prod_{\stackrel{s,k_z>0}{n \neq 0}}
e^{- \frac{k_B T}{m} \lambda_{s,\vec{k},n}
{\left|  J_{\vec{k},s,n} \right|}^{2} }
=
e^{- \frac{1}{2} \frac{1}{N}
 \sum_{s,k}
 {\left| \hat{q} \cdot \hat{\varepsilon}_{\vec{k},s} \right|}^{2}  
d_{\omega_{s,k}}(\tau -\sigma) q^2 } \nonumber \\
&=& e^{- \frac{1}{2} \frac{1}{N D} \sum_{s,k} d_{\omega_{s,k}}(\tau - \sigma) q^2 }
= e^{- \frac{1}{2} d_D (\tau - \sigma) q^2 }
\label{eqn:mean_exp_III}
\end{eqnarray}
where the component of frequency $\omega_{s,k}$ of the imaginary time diffusion
$d_D(\tau)$ is ($A_{\gamma}=A_{\gamma}(\omega_{s,k}), \Omega_{\gamma}= \Omega_{\gamma}(\omega_{s,k})$)
\begin{equation}
d_D(\tau) = \frac{1}{N D} \sum_{s,k}
d_{\omega_{s,k}}(\tau)  \nonumber 
\end{equation}
\begin{equation}
= \frac{1}{N D} \sum_{s,k} \frac{\hbar}{m} \sum_{\gamma}
\frac{A_{\gamma}}{\Omega^2_{\gamma}}
\frac{
\cosh\left( \beta \Omega_{\gamma} /2 \right) -
\cosh \left( \Omega_{\gamma} [\beta/2 -\tau] \right)}
{ \sinh \left( \beta \Omega_{\gamma} \right) } \label{eqn:d_tau_II}
\end{equation}

\section{APPENDIX D: MEAN ELECTRONIC FLUCTUATION}

The relation between the mean electronic fluctuation  and
the imaginary time diffusion $d_D (\tau)$ eq.(\ref{eqn:d_tau})  is
\begin{equation} \label{eqn:dif_sigma}
d_D \left( \tau \right)= \frac{2}{D}
\left[ <
{\left| \vec{u}(0) \right|}^2  > -
<  \vec{u}(\tau) \cdot \vec{u}(0) >
\right]
\end{equation}
comparing eq.(\ref{eqn:dif_sigma}) and 
eq.(\ref{eqn:d_tau_II}) for $d_D (\tau)$ and inserting
the DOS function, we have
\begin{equation} \label{eqn:sigma2}
\sigma^2_T = 
\frac{< {\left| \vec{u} \right|}^2  >}{D}  =
\int \!\!\! d \omega \rho(\omega)
\sum^{3}_{\gamma=1}
\frac{\hbar A_{\gamma}(\omega)}{2 m \Omega^2_{\gamma}(\omega)}
\coth \left( \frac{\beta \Omega_{\gamma}(\omega)}{2} \right) 
\end{equation}
If we fix $w=\omega_o$, we have $\Omega_3=\omega_o$ for one solution 
of the cubic polynomial eq.(\ref{eqn:pol3}) and by eq.(\ref{eqn:A_factors}) 
we have also $A_3=0$. The other two terms give 
\begin{equation} \label{eqn:del2_+_eq} 
{\left< u^2 \right>}_{+} = \!\!\!\!
\int \!\! d \omega \rho \left( \omega \right)
\frac{\Omega^2_1 - \omega^2_{LO}}{\Omega^2_1 - \Omega^2_2}
\frac{\hbar D}{2 m \Omega_1}
\coth
\left( \frac{\hbar \Omega_1}{2 k_B T} \right)  
\end{equation}
\begin{equation} \label{eqn:del2_-_eq}
{\left< u^2 \right>}_{-} =  \!\!\!\!
\int \!\! d \omega \rho \left( \omega \right)
\frac{\Omega^2_2 - \omega^2_{LO}}{\Omega^2_2 - \Omega^2_1 }
\frac{\hbar D}{2 m \Omega_2}
\coth
\left( \frac{\hbar \Omega_2}{2 k_B T} \right)  
\end{equation}
Notice that if we take a single Wigner frequency  being representative of the
electronic spectrum ($\rho (\omega)=\delta(\omega-\omega_W)$) we recover the
results of ref. \cite{Simone_meanfield}.

\section{APPENDIX E: POLARON RADIUS}

We now calculate the density-density correlation function 
of the eq.(\ref{eqn:C_self}) for the variational harmonic 
action $\mathcal{S}_T$.
We assume that the equilibrium position of
the reference electron $\imath=1$ is the origin.
With the same method to obtain eqs.(\ref{eqn:mean_exp_III}),
we performed the following Gaussian integrals 
for the density distribution $\rho_1 (\vec{r})$
\begin{equation}
{\left< \hat{\rho}_1 (\vec{r}) \right>}_{T} =
\int \!\! \frac{d^D q}{ {\left(2 \pi \right)}^D}
e^{\imag \vec{q} \vec{r}_e }
{\left< e^{\imag \vec{u}_1 \vec{q}} \right>}_{T} 
=
\frac{e^{- r^2 / 2 \sigma^2_T }}
{ {\left[2 \pi \sigma^2_T \right]}^{D/2} }
\end{equation}
and
\begin{equation} \label{eqn:2rela}
{\left<
e^{- \imag \vec{q} \vec{u}_1 }
e^{\imag \vec{q'} \left( \vec{u}_1 (\tau) - \vec{u}_1 \right) }
\right>}_{T}  
=
e^{- \frac{\sigma^2_T}{2} q^2} 
e^{- \frac{d(\tau)}{2} \vec{q'} \cdot \left[\vec{q'} + \vec{q} \right]}
\end{equation} 
Inserting eq.(\ref{eqn:2rela}) in  eq.(\ref{eqn:C_self}), we have 
the density-density correlation function in the imaginary-time 
for the $\imath=1$ electron
\begin{equation} \label{eqn:rho_rho}
{\left< \rho_1 (\vec{r})
\rho_1 (\vec{r'},\tau) \right>}_T = {\left< \hat{\rho}_1 (\vec{r}) \right>}_T
\frac{ e^{ -
\frac{{\left| \vec{r} - \vec{r'} +  \frac{ d(\tau)}{2 {\sigma}^{2}}  \vec{r}   \right|}^{2}}
{2\ell^2(\tau) } }
}
{ {\left(2 \pi \ell^2(\tau) \right) }^{D/2}}
\end{equation}
where
\begin{equation} \label{eqn:defll}  
\ell^2(\tau) = d_D(\tau) \left[ 1 - \frac{d_D(\tau)}{4 \sigma^2_T} \right]
\end{equation}
We notice that the function of eq.(\ref{eqn:rho_rho})
does not depend only on the relative distance
${\vec{r}}^{'} - {\vec{r}}$  but also
from the distance of electron from its
localization position in  the crystal.Then the eq.(\ref{eqn:C_self}) becomes 
\begin{equation} \label{eqn:C_self_T}
C^{self}_{1,T} =
\frac{1}{\epsbar} 
\int^{\beta}_{0} \!\!\! d \tau \frac{\omega_{LO}}{2} D_o(\tau)
\frac{
{\left< \rho_1 (\vec{r}) 
\rho_1 (\vec{r'},\tau) \right>}_{T}
}
{{\left< \rho_1 (\vec{r}) \right>}_{T}} 
\end{equation}
We assume $\vec{r} = 0$ (electron in its lattice point)
and then we obtain the variational 
radial induced charge density 
\begin{equation}\label{eqn:g_r_T}
g_T(r) = \frac{\pi \omega_{LO}}{2\epsbar} {\left( 2 r \right)}^{D-1}
\int^{\beta}_{0} \!\!\! d \tau  D_o(\tau)
\frac{ e^{ - r^2 / 2 \ell^2 (\tau)} }
{ {\left(2 \pi \ell^2(\tau) \right)}^{3/2} }
\end{equation}
by eq.(\ref{eqn:R_p}) we obtain the variational polaron radius 
\begin{equation} \label{eqn:R_p_T}
R_{p,T} =
{\left( \mbox{D} \; \frac{\omega_{LO}}{2} \int^\beta_0 \!\!\! d\tau D_o (\tau)
\ell^2(\tau)
\right)}^{1/2}
\end{equation}

\subsection{High density limit}

The characteristic length $\ell^2(\tau)$ defined in 
eq.(\ref{eqn:defll}) is expressed in term of $\tau$-dependent positional 
fluctuations $d_D(\tau)$, eq.(\ref{eqn:d_tau}), which is 
an integral of a function $d_{\omega_{s,k}} (\tau)$
weighted by the DOS $\rho (\omega)$ of the Wigner lattice,  
eq.(\ref{eqn:d_tau_II}).To have an estimate of this integral 
we replace  the integration by inserting an average  frequency 
in the function $d_{\omega_{s,k}}$. We choose  $\omega_P / \sqrt{\epsinf}$ 
because it is the typical frequency 
of the electronic fluctuation in the crystal
for the high density regime eq. (\ref{eqn:del2_+high_eq}). 
Moreover we consider the low temperature 
limit ($k_B T \ll \hbar \omega_{P} / \sqrt{\epsinf}$). 
Then from eq.(\ref{eqn:defll}) we get the following estimate for $\ell^2(\tau)$ 
\begin{equation} \label{eqn:ell2_high}
\ell^2 \left(\tau \right)  \simeq  
\frac{\hbar}{m \omega_P/ \sqrt{\epsinf}}
\left( 
1 - e^{- 2 \frac{\omega_P}{\sqrt{\epsinf}} \tau}
\right) 
\end{equation}
The characteristic time scale of electronic diffusion in imaginary time  is 
$\tau_{el} = {( \omega_P / \sqrt{\epsinf} )}^{-1}$. 
The rising-time is $\tau_{el} = 1/ (2 \omega_P /\sqrt{\epsinf})$.
Therefore we have  approximately
\begin{eqnarray*}
\ell^2 (\tau) &\simeq& \frac{\hbar}{m} \tau \quad (\tau \ll \tau_{el}) \\
\ell^2 (\tau) &\simeq&  
\frac{\hbar}{2 m \frac{\omega_P}{\sqrt{\epsinf}} }   \quad (\tau \gg \tau_{el})
\end{eqnarray*}
Now in the variational polaron radius $R_{p,T}$ 
of eq.(\ref{eqn:R_p_T}) another time 
scale appears  $\tau_{ph}=\omega_{LO}^{-1}$ 
but at high density $\tau_{ph}>>\tau_{el}$. 
Now we can separate the lowest time scale $\tau_{el}$ 
contribution  in the imaginary time integral 
so that we can approximate 
the integral in eq.(\ref{eqn:R_p_T}) as
\begin{eqnarray*}
\int^{\tau_{el}}_{0} \!\!\! d \tau  D_o(\tau)
\frac{ e^{ - \frac{r^2}{ 2 \ell^2 (\tau)} } }
{ {\left(2 \pi \ell^2(\tau) \right)}^{3/2} } &\simeq&
D_o(0)
\int^{\tau_{el}}_{0} \!\!\! d \tau \frac{ e^{ - \frac{r^2}{ 2 
\frac{\hbar}{m} \tau } } }
{ {\left(2 \pi \ell^2(\tau) \right)}^{3/2} } \\
& = & \frac{m}{2 \pi \hbar} \frac{1}{r}
\left(
1 - \mbox{erf} \sqrt{ \frac{r^2}{\left< u^2 \right>} }
\right) \\
\int^{\beta}_{\tau_{el}} \!\!\! d \tau  D_o(\tau)
\frac{ e^{ - 
\frac{r^2}{ 2 \ell^2 (\tau)}  
} }
{ {\left(2 \pi \ell^2(\tau) \right)}^{3/2} }
&\simeq&
\frac{ e^{ -  \frac{r^2}{ \frac{\hbar}{m \omega_P / \sqrt{\epsinf}  }} 
} }
{ {\left(2 \pi
\left< u^2 \right>
\right)
}^{3/2} }
\int^{\beta}_{\tau_{el}} \!\!\! d \tau  D_o(\tau) \\
&\simeq&
\frac{ e^{ - r^2 / 2 \left< u^2 \right>  } }
{ { \left(2 \pi
\left< u^2 \right> \right)}^{3/2} }.
\end{eqnarray*}
Collecting these results we get eq. (\ref{eqn:gr_approx}).


\begin{thebibliography}{99}
\bibitem{footnote_epsinf} 
The pure  electron gas limit of action
eq.(\ref{eqn:S_eff}) can be reached when  $\alpha=0$ and $\alpha_e$ is
finite, which {\it implies}  $\eta=1$. On the other hand,  for a given $\eta$,
$\alpha_e\rightarrow 0$  as $\alpha \rightarrow 0$. 
It should be noted, however,  that 
the limit of zero e-ph interaction is somehow unphysical in the presence of a given 
inverse polarizability parameter 
of the system:  as can be seen by the calculation of appendix B, when 
$\epszero \neq  \epsinf$, the  interaction with the
polarizable medium is \textit{necessary} to stabilize the crystal phase (static term).
\bibitem{footnote_gamma} In this sense the stabilization of the solid phase 
due to Fermi statistics is here considered not explicitly, since we consider 
distinguishable particles, but phenomenologically  via the fitted parameter $\gamma(\eta_q)$.
\bibitem{nota_m_pol} Notice however  that  this discussion does not involve the l.h.s. 
of eq.(\ref{eqn:del_L}) in which $\omega_P$ appears as a result of harmonic  approximation.
\bibitem{footnote_D_o} 
Notice that $\ell^2(\tau)$ and $d(\tau)$ are symmetric function 
respect to time $\beta / 2$ in the interval 
$0 < \tau < \beta$ as $D_o(\tau)$.
\end{thebibliography}
\end{document}